\documentstyle[12pt,ZJCITE]{article}
\newcommand{\be}{\begin{equation}}
\newcommand{\ee}{\end{equation}}
\newcommand{\bear}{\begin{eqnarray}}
\newcommand{\ear}{\end{eqnarray}}
\renewcommand{\theequation}{\arabic{section}.\arabic{equation}}
\newcommand{\GeV}{\mbox{$\;$GeV}}

\newcommand{\fm}{\rm fm}
\newcommand{\bm}[1]{\mbox{\bf #1}}
\renewcommand{\vec}{\bm}
\newcommand{\grgl}{\:\hbox to -0.2pt{\lower2.5pt\hbox{$\sim$}\hss}
           {\raise3pt\hbox{$>$}}\:}
\newcommand{\klgl}{\:\hbox to -0.2pt{\lower2.5pt\hbox{$\sim$}\hss}
           {\raise3pt\hbox{$<$}}\:}
\begin{document}
\begin{titlepage}
\begin{flushright}
HD--THEP--94--36
\end{flushright}
\quad\\
\vspace{1.8cm}
\begin{center}
{\bf\LARGE SOFT PHOTONS}\\
\bigskip
{\bf\LARGE  IN HADRON-HADRON COLLISIONS:}\\
\bigskip
{\bf\LARGE SYNCHROTRON RADIATION }\\
\bigskip
{\bf\LARGE FROM THE QCD VACUUM?}\\
\vspace{1cm}
G. W. Botz, P. Haberl, and O. Nachtmann\\
\bigskip
Institut  f\"ur Theoretische Physik\\
Universit\"at Heidelberg\\
Philosophenweg 16, D-69120 Heidelberg, FRG\\
\vspace{3cm}
{\bf Abstract:}\\
\parbox[t]{\textwidth}{We discuss the production of soft photons in
high
energy hadron-hadron collisions. We present a model where quarks and
antiquarks in the hadrons emit ``synchrotron light'' when being
deflected
by the chromomagnetic fields of the QCD vacuum, which we assume to
have a
nonperturbative structure. This gives a source of prompt soft photons
with
frequencies  $\omega\klgl $ 300 MeV  in the c.m. system of the
collision in
addition to hadronic bremsstrahlung. In comparing the frequency
spectrum and
rate of ``synchrotron'' photons to experimental results we find some
supporting evidence for their existence. We make an
exclusive--inclusive
connection argument to deduce from the ``synchrotron'' effect a
behaviour of
the neutron electric formfactor $G_E^n(Q^2)$ proportional to
$(Q^2)^{1/6}$
for $Q^2\klgl $ 20
fm$^{-2}$. We find this to be consistent with available data. In our
view,
soft photon production in high energy hadron-hadron and lepton-hadron
collisions as well as the
behaviour of electromagnetic hadron formfactors for low $Q^2$
are thus sensitive probes of the nonperturbative structure of the QCD
vacuum.
 }
\end{center}\end{titlepage}
\newpage
\section{Introduction}
\setcounter{equation}{0}
According to current theoretical ideas the vacuum state of quantum
chromodynamics (QCD) has a rich structure: it is characterized by
various condensates of quarks and gluons. We shall study here
possible
effects of the gluon condensate in high energy hadron-hadron
collisions:
the emission of soft prompt photons in addition to photons from
hadronic
bremsstrahlung. For frequency $\omega\to0$ the bremsstrahlung photons
must
dominate over photons from any other source due to Low's theorem
\cite{1} which, in essence is just a consequence of electromagnetic
gauge
invariance. The occurrence of additional soft prompt photons with
frequencies
$\omega\klgl $ 300 MeV was predicted in \cite{2}.
There the following physical picture was developed: When quarks ($q$)
and antiquarks $(\bar q)$ of a fast hadron move through the QCD
vacuum
they are subjected to the chromoelectric and chromomagnetic vacuum
fields.
The chromomagnetic Lorentz force will cause deflections of $q$ and
$\bar q$,
will make them wiggle around a bit. Since $q$ and $\bar q$ carry
electric
charge, they will then emit synchrotron radiation. For a single
hadron these
synchrotron photons are part of the photon cloud around the hadron.
But in
a high energy hadron-hadron, lepton-hadron, or photon-hadron
collision
the photons can be set free, manifesting themselves as prompt photons
with the characteristic energy spectrum of synchrotron radiation.

In this paper we will elaborate on the above idea. We will compare
the
predictions of our synchrotron effect with recent experimental
results
on soft photon production, especially the results of \cite{3}. We
will also
make a duality argument to relate the synchrotron effect to the
behaviour of
the electromagnetic formfactors of hadrons as function of the
momentum
transfer squared $Q^2$.

\section{Synchrotron Radiation from the \protect\\QCD Vacuum}
\setcounter{equation}{0}
The idea that there may be nonvanishing colour fields in the vacuum
of QCD has
a long history \cite{4-10} (for recent reviews cf. \cite{11,12}).
One important quantity characterizing the vacuum of QCD is the gluon
condensate
introduced in \cite{5,7}: There is ample phenomenological evidence
that the
product of two gluon field strengths at the same space-time point has
a
nonvanishing vacuum expectation value:
\be\label{2.1}
G_2:=\left\langle
0\left|\frac{g^2}{4\pi^2}G^a_{\mu\nu}(x)G^{a\mu\nu}(x)
\right|0
\right\rangle=
\frac{1}{4\pi^2}(0.95\pm0.45)\;{\rm GeV}^4.\ee
Here and in the following we always work in Minkowski space;
$G^a_{\mu\nu}(x)$
is the gluon field strength tensor with Lorentz indices $\mu,\nu$ and
colour
index $a$ $(a=1,...,8)$ and $g$ is the QCD coupling constant. All our
notation follows \cite{13}. The numerical value for $G_2$ is as
quoted
in \cite{12}. Considering the uncontracted product of two field
strength
tensors and using Lorentz and parity invariance one finds
\be\label{2.2}
\left\langle
0\left|\frac{g^2}{4\pi^2}G^a_{\mu\nu}(x)G^b_{\rho\sigma}(x)
\right|0\right\rangle
=\frac{1}{96}\delta^{ab}(g_{\mu\rho}g_{\nu\sigma}-
g_{\mu\sigma}g_{\nu\rho})
G_2.
\ee
For the chromoelectric and chromomagnetic fields this implies:
\bear\label{2.3}
&&\langle 0|g^2\bm B^a(x)\bm B^a(x)|0\rangle=\pi^2G_2\simeq(700\ {\rm
MeV})^4,\nonumber\\
&&\langle 0|g^2\bm E^a(x)\bm E^a(x)|0\rangle=-\pi^2G_2.\ear
Here one always assumes that the product of the field strengths is
normal
ordered with respect to the ``perturbative vacuum''. Thus (\ref{2.3})
should be interpreted in the following sense: in the physical vacuum
the chromomagnetic field fluctuates with bigger, the chromoelectric
field
with smaller amplitude than in the ``perturbative vacuum''.

The extension of this condensate idea to a correlation function of
gluon
fields at different space-time points was considered in
\cite{14,15,2,16,17}.
In this approach the vacuum is in essence
characterized by two numbers: the strength of the vacuum fields,
given
by $G_2$, and the correlation length $a$. This latter number is well
determined phenomenologically \cite{12,18} to be
\be\label{2.4}
a\simeq 0.35\ \fm.\ee

A physical picture of the QCD vacuum state incorporating the gluon
condensate $G_2$ and the finite correlation length $a$ is the domain
picture
as discussed in \cite{2}. In Euclidean space we can think of having
domains in the vacuum of linear size $\simeq a$ (cf.~Fig.~3 of
\cite{2}).
Inside one domain the colour fields are highly correlated. From one
domain to
the next the correlation is small. Of course, the orientation of the
colour
fields inside a domain and the domain sizes fluctuate. If we
translate this
picture naively to Minkowski space, we arrive at colour correlations
there being characterized by \underbar{invariant} distances of order
$a$.
Then the colour fields at the origin of Minkowski space, for
instance,
should be highly correlated with the fields in the region
\be\label{2.5}
|x^2|\klgl a^2\ee
(cf.~Fig.~1). Here it is important to remember that the gluon field
strengths
at point $x$ are to be compared with the gluon field strengths at the
origin
after
parallel transport on a straight line from $x$ to 0 (cf.
\cite{17,18}).

Let us introduce an observer in the ``femtouniverse'', i.e.\ an
observer
watching with a microscope of resolution much better than 1~fm a
hadron
which moves very fast in positive 3-direction. The observer will see
quarks
and gluons in the hadron. Consider for example a fast quark passing
through the
origin of the coordinate system in Fig.~1 on a nearly light-like
world line.
The above picture of the QCD vacuum implies that this fast quark
spends a long
time (from the point of view of the observer) in a highly correlated
colour
field.

In the following we will only consider ``light'' hadrons having no
``heavy''
valence quarks $c,b$. Then the hadron's radius $R\simeq 1\, \fm$ is
substantially larger than the correlation length $a$ (\ref{2.4}).

Let our observer now watch two quarks $q_1,q_2$ of the same hadron.
We want
to estimate the time interval over which $q_1$, moving in one domain,
can be
considered as moving uncorrelated with  $q_2$.

Let $E_h,E_1,E_2$ be the energies of the hadron and of $q_1,q_2$ in a
system
where they all move fast in positive 3-direction and let $m_q$ be the
quark
mass. For simplicity we consider the case where the transverse
velocities
of $q_1$ and $q_2$ are zero. The world lines of $q_{1,2}$ are then
\be\label{2.6}
y_i(t_i)\simeq\left(\begin{array}{c}
t_i\\\vec y_{iT}\\ y_i^3(0)+(1-\frac{\textstyle 1}{\textstyle
2\gamma_i^2})\,t_i\end{array}\right),\qquad (i=1,2),
\ee
where
\be\label{2.7}
\gamma_i=(1-\vec v_i^2)^{-1/2}=E_i/m_q\ee
are the Lorentz $\gamma$ factors for which we have assumed
\be\label{2.8}
\gamma_2\geq\gamma_1\gg1.\ee
We get now
\bear\label{2.9}
(\Delta y)^2&:=&\left(y_1(t_1)-y_2(t_2)\right)^2\nonumber\\
&\simeq&-2\left(y_1^3(0)-y_2^3(0)\right)(t_1-t_2)-(\vec y_{1T}-\vec
y_{2T})^2\nonumber\\
&&-\left(y^3_1(0)-y_2^3(0)\right)^2
+{\cal O}\left(\frac{1}{\gamma^2_i}\right).\ear
Typically $q_1$ and $q_2$ will have a transverse separation of order
$R$,
i.e.\ substantially larger than the correlation length $a$:
\be\label{2.10}
|\vec y_{1T}-\vec y_{2T}|={\cal O}(R).\ee
The longitudinal separation of $q_1$ and $q_2$ on the other hand will
be
Lorentz-contracted and can be estimated to be of order $R/\gamma_1$:
\be\label{2.11}
|y^3_1(0)-y_2^3(0)|\simeq R/\gamma_1.\ee
Thus we obtain from (\ref{2.9})--(\ref{2.11}), neglecting the terms
of
order $1/\gamma_i^2$:
\be\label{2.11a}
(t_1-t_2)\frac{R}{\gamma_1}\, c_1\simeq c_2R^2+\Delta y^2\ee
where $c_1,c_2$ are constants with $|c_1|,|c_2|$ of order 1. If we
require
now
\be\label{2.12}
|(\Delta y)^2|\klgl  a^2\ll R^2\ee
the solution of (\ref{2.11a}) is
\be\label{2.13}
t_1-t_2\simeq \frac{c_2}{c_1}\gamma_1R+\frac{\gamma_1}{c_1R}\Delta
y^2.
\ee
Setting here for an order of magnitude estimate $c_1=\pm 1$ and
$c_2=\pm 1$ we
get:
\be\label{2.13a}
|t_1-t_2|\simeq \gamma_1R(1\pm\frac{a^2}{R^2}).\ee
This means the following: The quark $q_2$ is in the same colour
domain
as $q_1$ at time $t_1$ only for the following intervals of the time
$t_2$:
\be\label{2.13b}
t_1-\gamma_1R(1+\frac{a^2}{R^2})\klgl t_2\klgl t_1-\gamma_1
R(1-\frac{a^2}{R^2})\ee
and
\be\label{2.13d}
t_1+\gamma_1R(1-\frac{a^2}{R^2})\klgl t_2\klgl t_1+\gamma_1
R(1+\frac{a^2}{R^2}).\ee
Conversely, for time separations
\be\label{2.13c}
|t_1-t_2|\klgl  \gamma_1R (1-\frac{a^2}{R^2})\simeq \gamma_1 R\ee
$q_1$ and $q_2$ are in \underbar{uncorrelated} domains. Note that the
time
interval (\ref{2.13c}) is large if $\gamma_1$ is large.

Introducing the energy fraction $x_1=E_1/E_h$ of hadron $h$ carried
by
$q_1$ we can write (\ref{2.13c}) as
\be\label{2.14}
|t_1-t_2|\klgl \frac{E_h}{m_q}x_1 R=:\Delta t(x_1).\ee
Our observer can interpret this result as follows. Two hard partons,
$x_2\geq x_1={\cal O}(1) $ are uncorrelated for a very long relative
time
separation $\Delta t(x_1)$. Short time
correlations only occur if $q_1$ is a ``wee'' parton, i.e.\ a parton
with
energy less than some fixed energy $E_{\rm wee}^{\rm max}\simeq 0.5$
GeV
$\simeq a^{-1}$, say. This
means for the energy fraction $x_{\rm wee}$ carried by a ``wee''
parton:
\be\label{2.15}
x_{\rm wee}\leq E_{\rm wee}^{\rm max}/E_h\simeq \frac{0.5\; {\rm
GeV}}{E_h}\simeq
\frac{1}{aE_h}.\ee
We find then that for all ``non-wee'' quarks, i.e.\ for all quarks
having
$x_2\geq x_1>E_{\rm wee}^{\rm max}/E_h$ the difference of times over
which they
travel
in an uncorrelated fashion satisfies
\be\label{2.16}
\Delta t(x_1)\grgl \frac{1}{m_q}E_{\rm wee}^{\rm
max}R\simeq\frac{1}{m_q}\frac{R}{a}
=:\Delta t_0.\ee
We think it is reasonable to put here for $m_q$ the current quark
masses
$m_u,m_d\simeq 5-10$ MeV \cite{19}, since we are concerned with
``non-wee''
partons. Then
\be\label{2.17}
\Delta t_0\simeq 50\ \fm,\ee
which is a very large time indeed for hadronic interactions.

Of course, in reality the quarks do not travel on parallel straight
lines.
In fact the effects which we discuss in this paper are due to the
bending
of the quark trajectories in the vacuum chromomagnetic fields. Thus
the
above exercise should only be taken as a rough estimate of how the
vacuum
structure with a \underbar{small} invariant correlation length $a$
(\ref{2.4})
can produce very \underbar{large} time and length scales $\Delta
t(x_1)\grgl\Delta
t_0$ (\ref{2.16}), (\ref{2.17}) for fast hadrons in Minkowski space.
We will
argue
below that this allows us to add the contributions to synchrotron
radiation from the partons in the hadrons incoherently.

At this point we can make a comment on the Drell-Yan process
\be\label{2.17a}
h_1+h_2\to V+X,\ee
where $h_{1,2}$ are the initial hadrons and $V$ is the produced
vector boson $(V=\gamma^*,W,Z)$. In \cite{20} possible
nonperturbative
effects related to the QCD vacuum structure were discussed for this
reaction. The basic process leading to (\ref{2.17a}) is the
annihilation
of a quark $q$ and an antiquark $\bar q$ with the production of the
vector
boson $V$:
\be\label{2.18}
q+\bar q\to V.\ee
Let the $q\bar q$-annihilation occur at the origin of the coordinate
system in
Fig.~2 and let $q(\bar q)$ be a parton of $h_1 (h_2)$. It is clear
that
here $q$ and $\bar q$ have travelled for a time $t \grgl \Delta t_0$
(\ref{2.17}) in a correlated colour background field. Thus there is
ample time
for a correlation of their spins to
occur as suggested in \cite{2,20}. Indeed, if we estimate the
build-up
time $\tau_{\rm pol}$ for the polarization due to gluon spin flip
synchrotron radiation from the analogous formula from electrodynamics
\cite{21}, we get for the quark $q$:
\be\label{2.19}
\tau_{\rm
pol}\simeq\frac{1}{\gamma_q^2}\frac{m_q^5}{\alpha_s(gB_c)^3}\,,\ee
where $gB_c$ is the effective chromomagnetic background field
strength times
the coupling constant and $\gamma_q$ is the Lorentz factor for the
quark.
For the antiquark a similar formula holds. Then certainly $\Delta
t_0\gg \tau
_{\rm pol}$ for large $\gamma_q,\gamma_{\bar q}$.  Of course, the
estimate
(\ref{2.19})  for $\tau_{\rm pol}$ applies to a static  $B$ field and
taking it
over
for the vacuum ``background'' fields is highly debatable. However,
what we may
conclude from  (\ref{2.19}) is that for $m_q\to 0$ the quark's  spin
motion and
corresponding  spin-flip synchrotron radiation are infrared
\underbar{sensitive}
effects related to the vacuum gluon fields. Thus even if we are
unable at
present to
calculate $\tau_{\rm pol}$ there is no reason for it to be  infinite
as one
would
expect for the perturbative vacuum where  $gB_c=0$.

We return now to photon ``synchrotron'' radiation in a hadronic
collision.
For definiteness we consider a nucleon-nucleon collision with the
production
of $n$ hadrons
\be\label{2.20}
N_1(p_1)+N_2(p_2)\to h_1(p_1')+...+h_n(p_n')\ee
in the overall c.m. system.

In \cite{2} we have estimated the number of synchrotron photons
emitted from
the nucleons in the initial state. Let $\vec k,\omega$ be the photon
momentum
and energy in the overall c.m. system. Then we obtain from (5.21),
(5.22) of
\cite{2} for the photons from $N_1(p_1)$
in (\ref{2.20}):
\be\label{2.21}
\omega\frac{d^3n_\gamma^{(N_1)}}{d^3k}=
\frac{\alpha}{\omega^{4/3}}(l_{\rm eff})
^{2/3}S^{(N_1)}(\hat{\bm{p}}_1\cdot\hat{\bm{k}})\ee
where $\hat{\bm{p}}_1=\vec p_1/|\vec p_1|,\ \hat{\bm{k}}=
\vec k/|\vec k|$ and
\be\label{2.22}
l_{\rm eff}=\frac{\sigma}{gB_c},\ee
\bear\label{2.23}
S^{(N_1)}(\hat{\bm{p}}_1\cdot\hat{\bm{k}})&=&\frac{1}{2\pi^2}
\frac{\Gamma^2(\frac{2}{3})
6^{1/3}}{2\ln(P/P_0)}\nonumber\\
&&\times \int^{\chi_0}_{-\chi_0}d\chi\,\Theta(\cos
\vartheta^*+\tanh\chi)(\cosh\chi)^{-2/3}
\nonumber\\
&&\times (1+\cos\vartheta^*
\tanh\chi)^{-4/3}\left(\frac{\cos\vartheta^*+\tanh\chi}
{\sin\vartheta^*}\right)^{2/3}\nonumber\\
&&\times \int^{\psi_m(\chi)}_0 \!\!\!d\psi\,\psi^{-1/3}
F_2^{N_1}\left(\frac{\psi}{\psi_m(\chi)}\right)
e^{-\frac{1}{2}\psi^2},\ear
\bear\label{2.24}
&&\cos \vartheta^*=\hat{\bm{p}}_1\cdot\hat{\bm{k}},\qquad
(0\leq\vartheta^*\leq\pi),\ear
\be\label{2.25}
\psi_m(\chi)=\frac{P\sin\vartheta^*}{\sigma|
\cos\vartheta^*+\tanh\chi|}
\frac{e^\chi}{\cosh\chi},\ee
\be\label{2.26}
P=p^0_1\simeq\frac{\sqrt s}{2} \, , \qquad \chi_0 = \ln(P/P_0).\ee
As explained in \cite{2}, we obtained these formulae by using the
synchrotron emission formulae for a constant chromomagnetic field in
a particular reference frame and then averaging incoherently over all
reference frames where both nucleons $N_1$ and $N_2$ move fast such
that
we can use the parton picture for them. In this way we introduced a
cutoff
parameter $P_0$, the minimal nucleon momentum for the parton picture
to
be applicable. We set $P_0=1.5$ GeV here. The quantity $\sigma$ is
the mean
transverse momentum of the quarks in the hadrons. We set
\be\label{2.27}
\sigma=300\ {\rm MeV}\ee
and we think that this number should be correct to $\pm 100$ MeV. The
quantity
$gB_c$ is, as in (\ref{2.19}),
the effective chromomagnetic background field strength times the
coupling
constant. Finally,~$l_{\rm eff}$ (\ref{2.22}) is the length or time a
fast
quark has to travel in a transverse background field of strength
$gB_c$ in
order to pick up a transverse momentum of magnitude $\sigma$
(cf.~Fig.~3).
$F_2^{N_1}$ is the usual structure function of the nucleon known from
deep
inelastic electron and muon nucleon scattering.

{}From the nucleon $N_2$ in the initial state of the reaction
(\ref{2.20})
we get of course a similar contribution of synchrotron photons:
\be\label{2.28}
\omega\frac{d^3 n_\gamma^{(N_2)}}{d^3
k}=\frac{\alpha}{\omega^{4/3}}(l_{\rm
eff})^
{2/3}S^{(N_2)}(\hat{\bm{p}}_2\cdot\hat{\bm{k}}),
\ee
where $S^{(N_2)}$ is obtained from (\ref{2.23}) with obvious
replacements:
$F_2^{N_1}\to F_2^{N_2},\;\vartheta^*\to$ $  )\hspace{-.9em}<  (\vec
k,\vec
p_2)$.

In the discussion above we have added the contributions of
synchrotron
photons from the partons in each nucleon $N_1,N_2$ incoherently and
we have
done the same for the total contributions from $N_1$ and $N_2$. This
needs some justification.

As we have seen in chapter 2, two partons of the same hadron travel
for most
of the time in different colour domains, i.e.\ the gluonic field
strengths
at their world lines are for most of the time uncorrelated. This
means
that their wiggling in the colour background fields is mostly
uncorrelated.
Thus interference terms from the emission of synchrotron photons by
two partons
will
tend to average to zero. This is elaborated on in Appendix A.

To add the contributions from $N_1$ and $N_2$ incoherently seems
quite
a harmless assumption to us, since for any given background
chromomagnetic
field the photons from the partons of $N_1$ and $N_2$ will populate
widely
different regions of phase space.

We will now argue that from the final state hadrons in (\ref{2.20})
we will
also get synchrotron photons. In the hadronic collision quarks,
antiquarks,
and gluons from the original hadrons will be deflected, and new
partons
will be created as well. These final state
partons will recombine to form hadrons. We can for instance think of
the
mechanisms discussed in the LUND model for particle production (for a
review
cf. \cite{22}). Shortly after the collision, however, these hadrons
did not
yet have time to ``dress'' themselves with soft synchrotron photons,
a process
which will, as we shall see, take a time of order 20~fm. Thus
initially we get ``undressed'' hadrons $h^{(0)}$. A state for a
physical
hadron $h$ can be expanded schematically as follows:
\be\label{2.29}
|h\rangle\simeq|h^{(0)}\rangle
+e|h^{(0)}\gamma\rangle,\ee
where the amplitude for $|h^{(0)}\gamma\rangle$ is due to our
synchrotron
effect
and gives the photon emission if $h$ is in the initial state.
Reversing this we get
\be\label{2.30}
|h^{(0)}\rangle\simeq|h\rangle-e|h\gamma\rangle,\ee
which we interpret as follows: The originally
produced ``naked'' hadron $h^{(0)}$ will lead with amplitude $\simeq
1$
to a physical final state hadron $h$ but also with some amplitude
of order $e$ to the emission of a ``synchrotron'' photon in addition.
We discuss the meaning of (\ref{2.29}) and (\ref{2.30}) on a more
formal level in Appendix B. The amplitudes for the photon emission in
(\ref{2.29})
and (\ref{2.30}) differ only by the overall sign. Thus we should get
a
contribution to synchrotron photons from each final state hadron $h$
with a spectrum given again by (\ref{2.21}) with suitable
replacements:
\be\label{2.31}
\omega\frac{d^3
n_\gamma^{(h)}}{d^3k}=\frac{\alpha}{\omega^{4/3}}(l_{\rm
eff})^{2/3}
S^{(h)}(\hat{\bm{p}}_h\cdot\hat{\bm{k}}).\ee
Here $S^{(h)}$ is as in (\ref{2.23}) with the replacements $F_2^{N_1}
\to F_2^{h},\;\vartheta^*\to\ )\hspace{-.9em}< (\vec k,\vec p_h)$.

We can now put everything together. For inelastic collisions
\be\label{2.32}
p(p_1)+N(p_2)\quad\to \quad{\rm hadrons}\quad h\ee
we predict the number of synchrotron photons per inelastic
collision in the overall c.m. system as
\be\label{2.33}
\omega\frac{d^3 n^{\rm (syn)}_\gamma}{d^3
k}=\frac{\alpha}{\omega^{4/3}}(l_{\rm
eff})^{2/3}
\Sigma(\hat{\bm{p}}_1\cdot\hat{\bm{k}}),\ee
where
\bear\label{2.34}
\Sigma(\hat{\bm{p}}_1\cdot\hat{\bm{k}})&=&
S^{(N_1)}(\hat{\bm{p}}_1\cdot
\hat{\bm{k}})+S^{(N_2)}
(-\hat{\bm{p}}_1\cdot\hat{\bm{k}})\nonumber\\
&&\!\!\!\!\!\!\!\!\!\!\!\!\!\!\!\!\!\!\!\!\!\!\!\!
+\sum_h\frac{1}{\sigma_{\rm inel}(N_1N_2)}\int\frac{d^3p_h}{E_h}
\left(E_h\frac{d^3\sigma}{d^3p_h}(N_1N_2\to hX)\right)\cdot
S^{(h)}(\hat{\bm{p}}_h
\cdot\hat{\bm{k}}).\ear
Here $h$ runs over all produced hadron species. We have in
(\ref{2.34}) added
the contributions of all hadrons from the initial and final state
incoherently. We think that this is justified since the time scale of
the
hadronic collision is of the order of 2--3~fm and thus very short
compared to the emission or ``dressing'' time of order 20~fm involved
in our synchrotron effect. Thus we treat the collision in a sort
of ``sudden approximation'' well known in quantum mechanics (cf.
e.g. \cite{22a}).

The photon spectrum from hadronic bremsstrahlung was calculated for
multiparticle
production in hadron-hadron collisions in \cite{51} on the basis of
Low's
theorem
\cite{1}. One finds for emission angles of the photon not too close
to the
collision  axis:
\be\label{2.40}
\omega\frac{d^3n_\gamma^{({\rm br})}}{d^3 k}
\propto\frac{1}{\omega^2[1-(\hat{\bm{p}}_1\cdot\hat{\bm{k}})^2]}.
\ee
{}From (\ref{2.33}) and (\ref{2.40}) we get for the ratio of
synchrotron to
bremsstrahlung photons:
\be\label{2.41}
\frac{\omega d^3n_\gamma^{({\rm syn})}}{d^3k}\bigg/ \frac{\omega d^3
n_\gamma^{({\rm
br})}}
{d^3 k}\propto \omega^{2/3}.
\ee
Thus our synchrotron effect vanishes relative to bremsstrahlung for
$\omega\to
0$
and Low's theorem is  respected. For very small $\omega$ the
synchrotron effect
vanishes relative to bremsstrahlung  faster than $\omega^{2/3}$
(\ref{2.41})
since
there the incoherent summation of the contributions of the different
partons is
no
longer justified. In Appendix A, below,  we estimate that this
incoherence
assumption should break down for
\be\label{2.42}
|\vec k_T|\klgl 3.5 \;{\rm MeV},\ee
where  $|\vec k_T|=|\omega\,\sin\vartheta^*|$.

\section{Comparison with Experiment}
\setcounter{equation}{0}

There are a number of experiments on prompt photon production in high
energy
hadron-hadron collisions. The earliest experiment  \cite{44} found a
$\gamma$-signal
compatible with bremsstrahlung from the hadrons. Subsequent
experiments
\cite{31-34}
 reported numbers of soft photons partly in great excess over the
expectations from hadronic bremsstrahlung. On the other hand  the
authors of
the
experiment \cite{3} give only upper limits on the possible presence
of
``anomalous''
soft photons and conclude that there cannot be too many of them.   An
anomalous
yield of prompt photons was also seen in muon-proton deep inelastic
scattering
\cite{361}. Here and in the following we will use the words
``anomalous yield
of
photons''  or simply ``anomalous photons'' for a yield exceeding the
theoretical
expectation from bremsstrahlung.

On the theoretical side in \cite{2} a prediction for the existence of
anomalous
soft
photons in hadron-hadron and lepton-hadron scattering as well as
jet production in
$e^+e^-$-annihilation was made.  This was \underbar{before} the
publication of
the
results of the first experiment \cite{31} which saw anomalously large
numbers
of
soft photons. Subsequently two other theoretical models for the
production of
anomalous soft photons were  proposed. The ``soft annihilation
model''
\cite{45,46}
which had previously been used to describe soft lepton pair
production
\cite{47,48}
and the ``glob'' model \cite{49,50}. We will comment on how to
distinguish
these
models below.

In this section we  will first compare our theoretical expressions
(\ref{2.33}),
(\ref{2.34}) to the experimental data from \cite{3}  and then make
some
comments
concerning the data from \cite{31-34}. In the experiment \cite{3}
soft
photons produced in
$p$--Be collisions at 450 GeV incident proton momentum were measured.
We have thus the collision
\be\label{3.1}
p(p_1)\ +\ N(p_2)\quad\to\quad {\rm hadrons}\ +\ \gamma(k)\ee
where $N$ stands for the average nucleon in the Be nucleus. The c.m.
energy of the $pN$ collision is
\be\label{3.2}
\sqrt s\ =\ 29.09\; {\rm GeV}.\ee
The final hadronic state contains in most of the cases two nucleons
plus pions. We will neglect other particle compositions of the
final state in the following.

In Fig.~4 we reproduce Fig.~5 of \cite{3} with the normalization
provided
to us by H. J. Specht. Here we use the variables in the overall c.m.
system:
\bear\label{3.3}
k_T&=&\left(\vec k^2-(\hat{\bm{p}}_1\cdot\vec
k)^2\right)^{1/2},\nonumber\\
y&=&-\ln\tan\left(\vartheta^*/2\right),\nonumber\\
\vartheta^*&=&\ )\hspace{-.9em}<  (\vec k,\vec p_1).\ear
Expressed in these variables our prediction for synchrotron photons
(\ref{2.33}), (\ref{2.34}) reads
\be\label{3.4}
\frac{d^2n_\gamma^{(\rm syn)}}{dk_Tdy}=\frac{2\pi\alpha(l_{\rm
eff})^{2/3}}{k_T^{1/3}}
(\sin\vartheta^*)^{4/3}\Sigma(\cos\vartheta^*).\ee
In the sum over the final state hadrons $h$ in (\ref{2.34}) we
include
2 nucleons plus pions, as mentioned above. Consider first the two
nucleons. These are ``leading'' particles and carry away a
substantial
fraction of the available energy in forward and backward direction,
respectively (cf. \cite{23}). We estimate their contribution to
$\Sigma$
to be approximately the same as for the initial state nucleons. Next
we consider the contribution to $\Sigma$ from $\pi^+,\pi^-,\pi^0$.
Due to
isospin and charge conjugation invariance the  structure functions
$F_2$ for
$\pi^+,\pi^-$ and $\pi^0$ are equal. At least approximately also the
inclusive cross sections for $\pi^+,\pi^-$ and $\pi^0$ are equal
\cite{23}. Making this assumption, we obtain from (\ref{2.34}) for
the
reaction (\ref{3.1})
\bear\label{3.5}
\Sigma(\cos\vartheta^*)&=&
2S^{(p)}(\hat{\bm{p}}_1\cdot\hat{\bm{k}})+2S^{(N)}
(-\hat{\bm{p}}_1\cdot\hat{\bm{k}})\nonumber\\
&&+3\frac{1}{\sigma_{\rm
inel}(pN)}\int\frac{d^3p_\pi}{E_\pi}\left(E_\pi
\frac{d^3\sigma}{d^3p_\pi}(pN\to\pi
X)\right)S^{(\pi)}(\hat{\bm{p}}_\pi\cdot\hat{\bm{k}}).\ear

To calculate $\Sigma$ numerically we used the following input. For
the structure functions $F_2^p,F_2^N$ entering in $S^{(p)},S^{(N)}$
(cf. (\ref{2.23})) we used the set $D_0^\prime$ from the
MRS analysis \cite{24}
with the
$Q^2$ value taken to be 4 GeV$^2$. In (6.37) of \cite{25} we have
estimated the ``equivalent'' $Q^2$ value of deep inelastic
lepton-hadron
scattering for a hadronic collision of c.m. energy $\sqrt s$ to be
\be\label{3.6}
Q^2\simeq\sqrt s\cdot0.05\; {\rm GeV}.\ee
For $\sqrt s\simeq 30$  GeV this gives $Q^2=1.5\; {\rm GeV}^2$. We
prefer
to use a slightly higher $Q^2$ value since structure functions at
$Q^2=1.5\;
{\rm GeV}^2$ are more likely to be contaminated by higher twist
effects
and thus their interpretation in terms of parton densities is more
questionable.
We use the set $D_0'$ which has structure functions $F_2(x)$
with a \underbar{finite} value at $x=0$. This we consider reasonable
for application in a hadronic reaction. We may cite here the duality
arguments between parton and produced hadron distributions put
forward
already in the pioneering work of Feynman \cite{26}: The rapidity
distribution
of the pions produced in a nucleon-nucleon collision is flat at zero
c.m.
rapidity. Assuming the same to hold for the partons in the
nucleons from which the pions originate leads to a behaviour
\be\label{3.6a}
F_2^N(x)\to const.\ {\rm for}\ x\to0.\ee
On the other hand, a significant rise of $F_2^p(x,Q^2)$ for
$x\klgl5.10^{-3}$
and $Q^2\grgl 9\; {\rm GeV}^2$ has been observed at HERA \cite{26a}.
This can hardly be relevant for our case since for a nucleon energy
in the c.m. system $E_N=15\; {\rm GeV}$,  values  of $x\leq5.10^{-3}$
correspond to
parton momenta $E_{\rm parton}\leq 45$ MeV. This is well in the
``wee''
region where we certainly should not use parton distributions fitting
high $Q^2$ data and evolving the structure  functions obtained in
this way to
$Q^2\simeq 4$ GeV$^2$ using the  leading twist theory only.
Of course, for very high energy hadron-hadron collisions the
situation will
be different. But this is beyond the scope of the present work.  We
also note
that
a recent fixed target muon-nucleon scattering experiment \cite{281}
presented
preliminary results for structure functions at low $x$ and low $Q^2$
supporting
rather the  $D_0^\prime$ type parton distributions over the singular
$D_-^\prime$
type of \cite{24}.

The structure function $F_2^\pi$ was taken from \cite{27}. We use
here
again $Q^2=4\; {\rm GeV}^2$. Our structure function input is shown in
Fig.~5.

For the inelastic $pN$ cross section we take
\be\label{3.7}
\sigma_{\rm inel}(pN)=33.2\ {\rm mb},\ee
where we assumed $\sigma_{\rm inel}(pp)=\sigma_{\rm inel}(pn)$ and
took
$\sigma_{\rm inel}(pp)$ for $\sqrt s=29$ GeV from the fit
on page III.77 of \cite{28}. For the mean pion multiplicity
at $\sqrt s=29$ GeV we took
\be\label{3.8}
\frac{1}{3}(n_{\pi^+}+n_{\pi^-}+n_{\pi^0})=3.74\ee
as deduced from the fits given in \cite{29}. For the pion inclusive
distribution we used  the parametrization given in \cite{30}, but
rescaled to give, when integrated, the pion multiplicity (\ref{3.8}):
\bear\label{3.9}
&&E\frac{d^3\sigma}{d^3p_\pi}(pN\to\pi X)=\frac
{AB^C}{(E_T+B)^C}f(y_\pi,p_T)\left\lbrace\begin{array}{ccc}
e^{-p_T}&{\rm for}& p_T<1\\
e^{-D(p_T-1)/\sqrt s-1}&{\rm for}& p_T>1\end{array}\right.\nonumber\\
&&(p_T\  {\rm in GeV}),\ear
where
\bear\label{3.10}
E_T&=&\sqrt{p^2_T+m^2_\pi},\nonumber\\
y_\pi&=&\frac{1}{2}\ln\frac{E+p_{||}}{E-p_{||}},\nonumber\\
E_{\rm max}&=&\frac{1}{2\sqrt s}(s+m^2_\pi-4m^2_p),\nonumber\\
p_{\rm max}&=&\sqrt{E^2_{\rm max}-m^2_\pi},\nonumber\\
p_{||\rm max}&=&\sqrt{p_{\rm max}^2-p_T^2},\nonumber\\
y_{\rm max}(p_T)&=&\ln\frac{E_{\rm max}+p_{||\rm
max}}{E_T},\nonumber\\
f(y_\pi,p_T)&=&\exp\left[-\frac{\alpha}{(y_{\rm
max}(p_T)-|y_\pi|)^\beta}\right],
\nonumber\\
A&=&\frac{3.74}{2.89}\cdot3.78\cdot10^{-24}{\rm cm}^2\;{\rm
GeV}^{-2},\nonumber\\
B&=&2{\GeV},\nonumber\\
C&=&12.3 ,\nonumber\\
D&=&23{\GeV},\nonumber\\
\alpha&=&5.13,\nonumber\\
\beta&=&0.38.\ear

Using these inputs we can calculate $S^{(p)},S^{(N)}$ and $\Sigma$.
For the
paramters
$P/\sigma$ and $P/P_0$ occurring in the integral (\ref{2.23}) we took
\be\label{3.11}
P/\sigma=50,\quad P/P_0=10\ee
for the $p$ and $N$ case and
\be\label{3.12}
P/\sigma=P/P_0=10\ee
for the pion case. We found that the results for photon emission
angles
$20^o\leq\vartheta^*\leq160^o$ were insensitive to variations in
these
parameters over a reasonable range. In Fig.~6 we show $2S^{(p)}(\cos
\vartheta^*)$ and
$\Sigma(\cos\vartheta^*)$ as function of $\vartheta^*$. As already
noted in
\cite{2}, we expect that
$\Sigma(\cos\vartheta^*)\propto(\sin\vartheta^*)^{-2/3}$ for most
$\vartheta^*$
 with the exception of angles very close to $\vartheta^*=0^o$
and $\vartheta^*=180^o$. This is confirmed by the numerical
calculations as
shown in Fig.~7.

Having determined $\Sigma$, our synchrotron radiation formula
(\ref{3.4})
contains as unknown parameter
only $l_{\rm eff}$. We did not try to make a ``best fit''
for $l_{\rm eff}$ from the data of Fig.~4.
Instead we looked at the result for the prompt
photon spectrum adding to the hadronic bremsstrahlung as given in
\cite{3}
the synchrotron photons (\ref{3.4}) with various values of $l_{\rm
eff}$.
We found that the data is quite consistent with $l_{\rm eff}$ in the
following
range
\be\label{3.13}
20 \;\fm\klgl l_{\rm eff}\klgl 40 \;\fm.
\ee
In Fig.~4 we show our results for $l_{\rm eff}=20$ fm and $l_{\rm
eff}=40$ fm
superimposed on the data.

We should be careful with the interpretation of our results. The
authors of
\cite{3} do \underbar{not} claim that there is an excess over the
hadronic bremsstrahlung in the photon spectrum. They interpret their
data as giving upper limits on the presence of additional sources
of direct photons at small transverse momentum.
So what we can conclude from our calculation is that the data is
consistent with the presence of ``synchrotron photons''
as additional source with $l_{\rm eff}
\klgl 40$ fm. Taking the error bars of the data
in Fig.~4 seriously we may even conclude that the data prefer
bremsstrahlung
plus synchrotron photons over bremsstrahlung photons only. It would
clearly
be very interesting to have more precise data which would allow to
draw more
definite conclusions.

At this point we should discuss the  data on soft photon production
in hadron-hadron collisions  from \cite{31-34}.
These experiments see a large excess of soft photons over the
bremsstrahlung level. For us theorists the experimental situation --
all data together -- is too confusing, so we will not attempt to
fit everything. Furthermore, comparison of theory with some published
experimental
results would require information on experimental cuts, detector
efficiencies
etc.
which is not available to us.
Anyway, our prediction of synchrotron photons is
given in (\ref{2.33}), (\ref{2.34}) and this should be for
experimentalists a target to shoot down.

When most of the present paper was written we received a preprint
\cite{321}
where
the soft annihilation model and the glob model are compared with
data. Here we
would
like to point out  that the dependence of the number $n_\gamma^{({\rm
an})}$
of
anomalous soft photons per event on the hadronic multiplicity $n_h$
provides a
way to
distinguish between these models and our synchrotron effect.
According to
\cite{321} the soft annihilation model gives $n_\gamma^{({\rm
an})}\propto
(n_h)^2$, in the glob model $n_\gamma^{({\rm an})}$ is independent of
$n_h$.
For our
synchrotron effect the multiplicity dependence of $n_\gamma^{({\rm
syn})}$ can
be
read off from (\ref{2.33}), (\ref{2.34}). In essence we have
\be\label{3.15}
n_\gamma^{({\rm
syn})}=\int\frac{d^3k}{\omega}\left(\omega\frac{d^3n_\gamma^{({\rm
syn})}}{d^3k}\right)\;\;\propto\;\; n_h +2\;,
\ee
where the 2 comes from the contribution of the initial state hadrons.

\section{Synchrotron Radiation and Electromagnetic \protect\\
Formfactors of Hadrons}
\setcounter{equation}{0}
We have argued in sect.~2 that the colour fields in  the vacuum
should
give a contribution to the virtual photon cloud of hadrons and
we made an estimate of the distribution of these photons using the
synchrotron radiation formulae. Consider now any reaction where a
quasi-real photon is emitted from a hadron with the hadron staying
intact and the photon interacting subsequently. In Fig.~8 we draw the
corresponding diagram for a nucleon $N$:
\be\label{4.1}
N(p)\to N(p')+\gamma(q).\ee
The flux of these quasi-real photons is well known. The first
calculations
in this context are due to Fermi, Weizs\"acker, and Williams
\cite{35}.
For us the relevant formula is given in eq.~(D.4) of \cite{36}.
Let $E$ be the energy of the initial nucleon, $G^N_E(Q^2)$ its
electric Sachs formfactor, and let $\omega$ and $q^2=-Q^2$ be the
energy and mass of the virtual photon. Then the distribution of
quasi-real photons in the fast moving nucleon is given by
\be\label{4.2}
dn_\gamma^{\rm
(excl)}=\frac{\alpha}{\pi}\frac{d\omega}{\omega}\frac{dQ^2}{Q^2}
\left[G^N_E(Q^2)\right]^2\ee
where we neglect terms of order $\omega/E$ and $Q^2/m^2_N$ and assume
\be\label{4.2a}
Q^2\gg Q^2_{\rm min}\simeq\frac{m^2_N\omega^2}{E^2}.\ee
We call (\ref{4.2}) the exclusive flux since the nucleon stays
intact.
Now we want to translate (\ref{4.2}) into a distribution in
$\omega$ and the angle $\vartheta^*$ of emission of the $\gamma$
(cf. Fig.~8).  We have from transverse momentum balance:
\bear\label{4.2b}
|\vec q|\sin\vartheta^*&=&|\vec p'|\sin\theta,\nonumber\\
|\vec q|^2 \sin^2\vartheta^*&=&|\vec
p'|^2(1-\cos\theta)(1+\cos\theta),\ear
Furthermore we have:
\be\label{4.3}
Q^2=2(EE'-m^2_N-|\vec p||\vec p'|\cos\theta),\ee
where $E=p^0,E'={p'}^0=E-\omega$.
{}From this we get
\bear\label{4.4}
\sin^2\vartheta^*&=&\frac{1}{\omega^2+Q^2}\frac{1}{|\vec p|^2}
\left(\frac{1}{2}Q^2-m^2_N\frac{\omega^2}{EE'+|\vec p||\vec
p'|-m^2_N}
\right)\nonumber\\
&&\times (EE'+|\vec p||\vec p'|-m^2_N-\frac{1}{2}Q^2).\ear
For
\be\label{4.4a}
E,E'\gg m_N,\omega \quad{\rm and}\quad \omega^2\gg Q^2\ee
this reduces to
\be\label{4.5}
\sin^2\vartheta^*\simeq\frac{Q^2}{\omega^2}.\ee
Inserting this in (\ref{4.2}) gives
\be\label{4.6}
dn_\gamma^{(\rm excl)}\simeq\frac{2\alpha}{\pi}\frac{d\omega}{\omega}
\frac{d\vartheta^*}{\sin\vartheta^*}\left[G^N_E
(\omega^2\sin^2\vartheta^*)\right]^2\ee
The range in $\vartheta^*$ for which (\ref{4.6}) is valid is obtained
from
(\ref{4.2a}) and (\ref{4.4a}), (\ref{4.5}) as
\be\label{4.6a}
\left(\frac{m_N}{E}\right)^2\ll\sin^2\vartheta^*\ll1.\ee

We will now make an ``exclusive-inclusive connection'' argument. In
(\ref{2.21}) we have calculated the total spectrum of synchrotron
photons from a nucleon in an \underbar{inelastic} collision:
\be\label{4.7}
dn_\gamma^{(\rm incl)}=
2 \pi\alpha\frac{d\omega}{\omega}d\vartheta^*\sin\vartheta^*
\omega^{2/3}(l_{\rm eff})^{2/3}S^{(N)}(\cos\vartheta^*).\ee
We call this now the inclusive spectrum, since it can be thought of
as
the integral result of the synchrotron effect from the diagram of
Fig.~8
and the corresponding ones, where instead of the outgoing nucleon
$N(p')$ we allow an arbitrary hadronic state $X$ to occur:
\be\label{4.8}
N(p)\to X(p')+\gamma(q).\ee
The exclusive channel (\ref{4.1}) is part of the inclusive sum and we
will
now require that it contains also a ``synchrotron'' piece, leading to
the
same power of $\omega$ as in (\ref{4.7}). We can achieve this by
requiring
a term in the formfactor $G_E^N(Q^2)$ being proportional to
$(Q^2)^{1/6}$.
Then, indeed, we find from (\ref{4.6}):
\be\label{4.9}
dn_\gamma^{(\rm excl)}\propto
\frac{2\alpha}{\pi}\frac{d\omega}{\omega}
d\vartheta^*\sin\vartheta^*\omega^{2/3}(\sin\vartheta^*)^{-4/3}.\ee
The exclusive distribution (\ref{4.9}) drops faster in angle than the
inclusive one (\ref{4.7}), where for $\vartheta^*$ satisfying
(\ref{4.6a})
$S^{(N)}(\cos\vartheta^*)\propto(\sin\vartheta^*)^{-2/3}$. This is
quite
reasonable
physically.

Thus we arrive at the following conclusion: The proton formfactor
$G_E^p$
should contain in addition to a ``normal'' piece connected with the
total
charge and the hadronic bremsstrahlung in inelastic collisions a
piece $\propto (Q^2)^{1/6}$ connected with ``synchrotron'' radiation
from
the QCD vacuum. In our exclusive-inclusive connection argument we
have
neglected any interference between these two pieces since this can be
expected to be very small in an inclusive sum. For the neutron
which has total charge zero we would expect the ``normal'' piece in
$G_E^n$ to be quite small and the ``anomalous'' piece to be quite
important for not too large $Q^2$. Thus the neutron electric
formfactor
should be an interesting quantity to look for ``anomalous''
effects $\propto(Q^2)^{1/6}$.

In Fig.~9 we show the data on the electric formfactor of the neutron
from
\cite{37,38}. We superimpose the curve
\be\label{4.10}
G^n_{(\rm syn)}(Q^2)=3.6\cdot
10^{-2}\left(\frac{Q^2}{5\,\fm^{-2}}\right)
^{1/6}\ee
which is normalized to the data at $Q^2=5\,\fm^{-2}$. We see that
except in
the very low $Q^2$ region we get a decent description of the data.
For
$Q^2\to0$ (\ref{4.10}) has to break down since $G_E^n(Q^2)$ is
regular
at $Q^2=0$. Indeed one knows the slope of $G_E^n(Q^2)$ for $Q^2=0$
from
the scattering of thermal neutrons on electrons (\cite{39} and
references
cited therein):
\be\label{4.11}
\frac{dG_E^n(Q^2)}{dQ^2}\Biggl|_{Q^2=0}=0.019\ \fm^2.\ee
We see from Fig.~9 that the behaviour of $G_E^n(Q^2)$ has to change
rather quickly as we go away from $Q^2=0$. In Fig.~10 we plot the
formfactor
in a double logarithmic diagram. A behaviour proportional to
$(Q^2)^{1/6}$
would lead to a straight line with slope 1/6. We see again that
within
the errors this is consistent with the data in the range
\be\label{4.12}
0.5\ {\fm}^{-2}\leq Q^2\leq 20\ {\fm}^{-2}.\ee
Note that the square of the inverse correlation length $a$
(\ref{2.4})
gives
\be\label{4.13}
a^{-2}=(0.35\ \fm)^{-2}=8.16\ \fm^{-2}.\ee
Thus it should not be surprising if we see effects of the QCD
vacuum structure in the $Q^2$ range (\ref{4.12}). Translating the
findings
(\ref{4.10}), (\ref{4.12}) into coordinate space we see that
according
to our results the charge density $\rho(r)$ and the electromagnetic
potential $V(r)$ of the neutron should behave as
\bear\label{4.14}
\rho(r)&\propto&r^{-10/3},\nonumber\\
V(r)&\propto&r^{-4/3}\ear
in the following range for the radius $r$:
\be\label{4.15}
1.5\ {\fm} \grgl  r \grgl 0.2\ {\fm}.
\ee

What about the  electric formfactor of the proton,  $G^p_E(Q^2)$? A
piece
proportional to $(Q^2)^{1/6}$ in  $G_E^p(Q^2)$  should produce a
rapid
variation
with $Q^2$ when going from $Q^2=0$ to $Q^2\simeq 2\;\fm^{-2}$ (cf.
Fig.~9).
This
should be difficult to fit with the usual  dipole formula:
\be\label{4.19a}
G_{\rm Dipole}(Q^2)=\frac{1}{[1+Q^2/m^2_D]^2}\, ,\ee
where
\be\label{4.19b}
m^2_D=0.71 {\rm GeV}^2=18.23 \;\fm^{-2}.
\ee
Deviations of $G^p_E(Q^2)$ from the dipole formula (\ref{4.19a}) have
indeed
been
reported for $Q^2\klgl 5 \;\fm^{-2}$ (cf. \cite{411} and references
cited
therein).
We leave quantitative fits of these effects with our ansatz to
further work.

If we apply the same exclusive-inclusive argument used above for the
nucleons
to the
electromagnetic formfactor of the pion we seem to run into a problem:
The
electromagnetic formfactor of the $\pi^0$ is strictly zero due to
current
conservation and  CPT invariance. On the other hand, the inclusive
synchrotron
photon radiation from a $\pi^0$ is nonzero
since the structure function $F_2^\pi(x)$ is nonzero (cf. the
analogue of
(\ref{2.23}) for a $\pi^0$). We can resolve this paradox in a way
suggested a
long
time ago
for the analogous case of deep inelastic scattering by assuming that
the
transition formfactor $\pi\to\rho$ in
\be\label{4.16}
\pi^0(p)\to\rho^0(p')+\gamma(q)\ee
in reactions analogous to those of Fig.~8 has a piece dual to
synchrotron
radiation. Thus we predict the presence of ``anomalous'' terms
$\propto
(Q^2)^{1/6}$ in the formfactors for (\ref{4.16}). The $\pi^0$
formfactor
being zero and thus having certainly no ``anomalous'' piece it would
seem natural to us that also the charged pion formfactor has no
``anomalous''
piece.

\section{Conclusions}
\setcounter{equation}{0}
In this article we have argued that the nontrivial vacuum structure
in QCD should lead to the emission of soft photons in addition
to hadronic bremsstrahlung in hadron-hadron collisions. We compared
our
theoretical results with the data from proton-Beryllium scattering at
450 GeV
incident proton momentum \cite{3}. We found that the data does indeed
allow for a contribution from ``synchrotron'' photons. We obtained a
decent description of the data  adding the photon spectra from
hadronic
bremsstrahlung and from our ``synchrotron'' effect (\ref{3.4}) with
an
effective
length
$l_{\rm eff}\simeq 20\ \fm$. The physical meaning of $l_{\rm eff}$ is
explained
in
Fig.~3. Inserting this in (\ref{2.22}) and using for the mean
transverse
momentum of quarks in a hadron $\sigma=300$ MeV (cf. (\ref{2.27}))
leads
to an effective field strength
\be\label{5.1} gB_c=\frac{\sigma}{l_{\rm eff}}\simeq(55\ {\rm
MeV})^2.\ee
This is much smaller than the vacuum field strength (\ref{2.3})
deduced from the QCD sum rules.
We must conclude that quarks in a \underbar{fast moving}
hadron see only a \underbar{shielded} vacuum chromomagnetic
field. This shielding is presumably done by the gluons in the
hadrons.
Indeed, we know from deep inelastic scattering that roughly 50 \% of
the
momentum of a fast hadron is carried by gluons. We find here a vital
role to play for these gluons: shielding the vacuum colour fields.
Indeed
if we insert the vacuum field strength (\ref{2.3}) and $\sigma=300$
MeV in
(\ref{2.22}) we get only a tiny $l_{\rm eff}=0.12\ \fm$! A wiggling
around of quarks in a fast hadron over such tiny length scales is
quite
contrary to all the physical picture which emerges from deep
inelastic
scattering. It would, for instance, lead to large violations of
the Callen-Gross relation \cite{41} which are not observed in
experiment.

We have then argued, by making an exclusive-inclusive connection,
that
the nucleon electromagnetic formfactors should contain terms
proportional
to $(Q^2)^{1/6}$ for $0<Q^2\klgl 1{\GeV}
^2$. We found that the data for the neutron electric formfactor shows
a behaviour quite consistent with this. Applying the same arguments
to
the pion formfactor we concluded that the electromagnetic formfactors
of $\pi^\pm$ should \underbar{not} contain such a $(Q^2)^{1/6}$
piece,
but the $\pi\to\rho$ transition formfactor should.

Clearly the ideas presented in this article are rather speculative.
However,
we can now cite at least three ``anomalous'' effects for which we
give
a common interpretation in terms of the QCD vacuum structure:

(1) the spin effects in the Drell-Yan process \cite{20},

(2) soft photon production in hadronic collisions,

(3) the $Q^2$ behaviour of the neutron electric formfactor.

We think it should be a worthwhile goal for experimentalists to
support or
disprove the above theoretical ideas with more and more precise data.

Finally, let us make some general comments and pose some questions on
anomalous
prompt soft photons in hadron-hadron and lepton-hadron scattering.

(1) The most important point is, of course, to establish or disprove
their
existence experimentally. A good place to check the presence
of anomalous photons are reactions where only neutral hadrons
participate, e.g.
\bear\label{5.2}
K_L+n&\to& K_L+n+\gamma,\nonumber\\
\Lambda+n&\to&\Lambda +n+\gamma.
\ear
Then bremsstrahlung is highly suppressed since it is only due to
magnetic
moments of the neutral hadrons and  ``anomalous'' photons should then
be a
clear signal. Experimental investigation of the reactions (\ref{5.2})
is not
totally hopeless. One could use $K_L$-- or $\Lambda$--deuteron
scattering with
the proton as spectator. Bremsstrahlung photons from this soft proton
should be
reliably calculable.

(2) Do the anomalous photons come from the particles in the initial
state,
in the final state or from both the initial and final state particles
 -- or do they come from some other mechanism?
The answer should be given by the
dependence of the photon yield on the hadronic multiplicity in the
collision.
It should also be interesting to look for correlations between
hadronic energy
flow and the photon emission angles. A ``thermal'' production
mechanism as
proposed e.g. in the ``glob'' model should produce  photons in a more
isotropic
way than  the synchrotron mechanism discussed here where the photons
follow
more or less the initial and final hadrons' directions.

(3) Do the anomalous photons come from the hadrons or from the
quarks in the
hadrons, i.e. do the quarks in the hadrons act coherently or
incoherently? It
should be possible to answer this question  by comparing reactions
where
hadrons with different charge but similar structure functions $F_2$
participate. Examples are
\bear\label{5.3}
\Lambda  +p&\to& {\rm hadrons}\,,\nonumber\\
\Sigma^++p&\to&{\rm hadrons}\,;
\ear
or
\bear\label{5.4}
p+p&\to&p+p+\mbox{\rm mostly neutral pions}\,,\nonumber\\
p+p&\to& p+p+\mbox{\rm mostly charged pions}\,.
\ear

(4) The frequency, angular, and $s$-dependence of the anomalous
photons should
be quite revealing. In this paper we have discussed in detail the
frequency and
angular dependence predicted in our model.  The $s$-dependence enters
in two
ways (cf. (\ref{2.34})): Through the hadronic  one particle inclusive
distributions  and through the synchrotron formula (\ref{2.21}),
(\ref{2.23})
and its  analogues for the final state hadrons. There we have to use
the
structure function $F_2$ at some effective $Q^2$. Taking the latter
to increase
with $\sqrt{s}$, as seems quite reasonable, we come sooner or later
into the
region  $Q^2\grgl 10$ GeV$^2$ where HERA data \cite{26a} show a large
increase
of $F_2$ for $x\to 0$. Taking for instance the estimate (\ref{3.6})
for the
effective $Q^2$ we find $Q^2\simeq 100$ GeV$^2$ for $\sqrt{s}=2000$
GeV, i.e.
at Tevatron energies. Thus our model -- as well as all models of soft
photon
production where the quarks in hadrons act incoherently -- will
predict a
substantial increase of the photon yield with c.m. energy $\sqrt{s}$
due to the
increase of the quark densities measured by $F_2$.

(5) The production of soft lepton pairs, $e^+e^-$ and $\mu^+\mu^-$ is
most
probably a related phenomenon since the lepton pairs surely come from
virtual
photons $\gamma^*$. There one has the possibility to investigate the
polarization of the virtual photons $\gamma^*$ by analysing the
angular
distributions of the lepton pairs in the $\gamma^*$ rest system.
Polarization
phenomena could lead to sensitive  tests of theoretical models.

We hope that some of our remarks above may be useful for the
discussion and
clarification of the intriguing  phenomenon of ``anomalous photons''.
\bigskip
\section*{Acknowledgments}
The authors would like to thank H. G. Dosch, B. R. French, D. Gromes,
G.
Ingelman,  P. V. Landshoff, F. Lenz, H. Leutwyler,
P. Lichard, M. Neubert, H. J. Specht, M.
Spyropoulou-Stassinaki, Th. Walcher
 for useful discussions and correspondence. Special
thanks are due to H. J. Specht for providing the normalization
of the  data of ref. \cite{3} on prompt photons and to P. V.
Landshoff and H.
G.
Dosch  for reading the manuscript, part of which was written while
one of the
authors
(O. N.) was visiting DAMTP of Cambridge University. This author would
like to
thank
P. V. Landshoff for the hospitality extended to him there and the
DAAD for supporting this visit financially under project Nr.
313-ARC-VIII-vo/scu.

\section*{Appendix A}
\renewcommand{\theequation}{A.\arabic{equation}}
\setcounter{equation}{0}
In this appendix we discuss the emission of ``synchrotron'' photons
by two
quarks of a hadron  and estimate the frequency range over which the
emission from two quarks can be considered as being incoherent.

Consider two quarks $q_{1,2}$ of a fast hadron as in Sect.~2,
eq.~(\ref{2.6})
ff. To calculate their synchrotron emission we treat the quarks as
classical
charged point particles moving in the background  chromomagnetic
vacuum field
(cf.~\cite{2}). Let the world lines  of $q_1$ and $q_2$ be
parametrized by the
times $t_i\;$:
\be\label{A.1}
z_i(t_i)=\left(\begin{array}{c}
t_i\\
\vec z_i(t_i)\end{array}\right)\;,\quad
i=1,2 \;.\ee
Consider first quark $q_1$ which moves in 3-space along the curve
${\cal C}_1$
\be\label{A.2}
{\cal C}_1:t_1\to\vec z_1(t_1).\ee
For all times $t_1$ we can  construct the tangent vector
${\cal{T}}_1$ and the vector ${\cal N}_1$ pointing from $\vec
z_1(t_1)$ to the
center of the instantaneous curvature circle. The standard formulae
of
differential geometry (cf. e.g. \cite{421}) give:
\bear\label{A.3}
{\cal{T}}_1(t_1)&=&\frac{\dot{\vec z}_1(t_1)}{|\dot{\vec
z}_1(t_1)|},\nonumber\\
{\cal N}_1(t_1)&=&\frac{1}{|\dot{\vec z}_1(t_1)|}
\frac{d{\cal{T}}_1(t_1)}{dt_1}.\ear
The radius of curvature of ${\cal C}_1$ at the point corresponding to
the time
$t_1$ is then
\be\label{A.4}
\rho_1(t_1)=(|{\cal N}_1(t_1)|)^{-1}.\ee
For a quark $q_1$ with Lorentz factor $\gamma_1(t_1)$ we can define
the
instantaneous cyclotron frequency and effective colour magnetic field
by
\bear
\omega_c^1(t_1)&:=&\frac{\sqrt{(\gamma_1(t_1))^2-1}}{\rho_1(t_1)},
\label{A.5}\\
g\,B_c^1(t_1)&:=&m_q\,\omega_c^1(t_1).\label{A.6}\ear
The next step is to construct the analogous quantities for quark
$q_2$.

The amplitude for emission of a photon of wave vector $\vec k$,
frequency
$\omega=|\vec k|$ and polarization vector  {\boldmath{$\varepsilon$}}
$(\mbox{\boldmath{$\varepsilon$}} \cdot\vec k=0)$ by the two quarks
is given by
(cf. e.g. \cite{422}):
\be\label{A.7}
A(\vec{k},\mbox{\boldmath{$\varepsilon$}})=ie \
\mbox{\boldmath{$\varepsilon$}}^*\vec{Z}(\vec{k}),\ee
where
\bear\label{A.8}
\vec Z(\vec k)&=&\sum^2_{i=1}\vec Z_i(\vec k),\nonumber\\
\vec Z_i(\vec k)&=&\int^\infty_{-\infty} dt_i\frac{d\vec
z_i}{dt_i}(t_i)e^{i(\omega t_i-\vec k\vec z_i(t_i))}\;,\quad
i=1,2\;.\ear
The synchrotron formulae are obtained by expanding the integrand in
(\ref{A.8})
around the time $t_i$ where
\be\label{A.9}
\vec k\cdot{\cal N}_i(t_i)=0.\ee
Let $t_{i,0}$ be the times where (\ref{A.9}) is satisfied and assume
that there
is just one solution of (\ref{A.9}) for $i=1$ and $i=2$. Set
furthermore
\bear\label{A.10}
\cos\vartheta_i&:=&\hat {\bm k}\cdot {\cal{T}}_i(t_{i,0}),\nonumber\\
\eta_i&:=&\frac{(t_i-t_{i,0})\omega_c^{i,0}}{(1+\gamma^2_{i,0}
\vartheta^2_i)^{1/2}}\nonumber\\
\xi_i&:=&\frac{\omega}{3\gamma^2_{i,0}
\omega_c^{i,0}}(1+\gamma^2_{i,0}
\vartheta_i^2)^{3/2}\;,\ear
where an index 0 indicates that the quantities are to be taken at
$t_i=t_{i,0}$.
We get then (cf. \cite{422}, \cite{21}):
\bear\label{A.11}
\vec Z_i(\vec k)&=&\exp(i\omega t_{i,0}-i\vec k\vec z_{i,0})
\int^\infty_{-\infty}d\eta_i
\frac{(1+\gamma_{i,0}^2\vartheta_i^2)^{1/2}}
{\omega_c^{i,0}}\nonumber\\
&&\times\left\{\dot{\vec
z}_{i,0}+\frac{(1+\gamma^2_{i,0}
\vartheta_i^2)^{1/2}}{\omega_c^{i,0}}\eta_i
\ddot{\vec z}_{i,0}\right\}\exp\left[i
\frac{3}{2}\xi_i(\eta_i+\frac{1}{3}\eta_i^3)\right],\\
\label{A.12}
\vec Z_i(\vec k)&=&
\frac{2}{\sqrt3}
\frac{(1+\gamma^2_{i,0}\vartheta^2_i)^{1/2}}{\omega_c^{i,0}}
\exp(i\omega t_{i,0}-i\vec k\vec z_{i,0})\nonumber\\
&&\times\left\{  \dot{\vec z}_{i,0}
K_{1/3}(\xi_i)+i
\frac{(1+\gamma^2_{i,0}\vartheta^2_i)^{1/2}}{\omega_c^{i,0}}
\ddot{\vec z}_{i,0}
K_{2/3}(\xi_i) \right\},\ear
where $K_{1/3,2/3}$ are modified Bessel functions.

The probability of photon emission is given by
\bear\label{A.13}
\omega\frac{d^3 n}{d^3k}&=&
\frac{\alpha}{4\pi^2}\sum_{\gamma{\rm -pol.}}
|\mbox{\boldmath{$\varepsilon$}}^*\cdot
\vec Z(\vec k)|^2\nonumber\\
&=&\frac{\alpha}{4\pi^2}\left(|\vec Z(\vec k)|^2-|\hat{\bm k}\cdot
\vec Z(\vec
k)|^2\right).\ear
The angular width of the photon distribution is estimated by
$\xi_i\klgl 1$,
since the Bessel functions $K_{1/3,3/2}(\xi_i)$ drop exponentially
for
$\xi_i>1$.

Suppressing here indices $i,0$ we get from (\ref{A.10}):
\bear\label{A.14} \xi&\klgl& 1\nonumber\\
\Longrightarrow\;\;
\vartheta^2&\klgl&\left(\frac{3}{\gamma}
\frac{\omega_c}{\omega}\right)^{2/3}=:\vartheta^2_{\rm max}\,.\ear
As to be expected, the emission occurs for
small $\vartheta$, i.e. in a narrow cone around the velocity vectors
$\dot{\vec z}_{i,0}$ for large $\gamma$. According to (\ref{A.11})
the
radiation is obtained from small time intervals $\Delta t$ around the
times
$t_{i,0}\,$.  We can estimate $\Delta t$ from the condition
$|\eta_i|\klgl 1$.
This gives, taking into account (\ref{A.14}):
\bear\label{A.15}
&&\frac{\Delta t\omega_c}{(1+\gamma^2\vartheta^2)^{1/2}}\simeq
1,\nonumber\\
\Longrightarrow&&\Delta t
\omega_c\left(3\gamma^2\,\frac{\omega_c}{\omega}\right)^{-1/3}\simeq
1,\nonumber\\
\Longrightarrow&&\Delta
t\;\simeq\;(3\gamma^2)^{1/3}\omega_c^{-2/3}\omega^{-1/3}.\ear
Another way to arrive at (\ref{A.15}) is to calculate the time
$\Delta t'$ over
which the velocity vector of the quark moving approximately on a
circle of
radius $\rho$ (\ref{A.4}) points inside a cone of angle
$\vartheta\leq\vartheta_{\rm max}$ around $\vec k$. Indeed, we find
from
(\ref{A.4}), (\ref{A.5}) and (\ref{A.14})
\be\label{A.16}
\Delta t'\simeq\rho\cdot\vartheta_{\rm
max}=\frac{\gamma}{\omega_c}\left(
\frac{3}{\gamma}\,
\frac{\omega_c}{\omega}\right)^{1/3}\equiv\Delta t.\ee

Consider first quarks being almost ``wee''. We have then from
(\ref{2.15})
\be\label{A.17}
\gamma=E^{\rm max}_{\rm wee}/m_q=\frac{1}{a\,m_q},\ee
\bear\label{A.18}
\Delta
t&\simeq&\left(\frac{1}{a\,m_q}\right)^{2/3}
\left(\frac{gB_c}{m_q}\right)^{-2/3}\omega^{-1/3}\nonumber\\
&=&\left(\frac{1}{a\,gB_c}\right)^{2/3}\omega^{-1/3}\nonumber\\
&=&\left(\frac{l_{\rm eff}}{a\,\sigma}\right)^{2/3}\omega^{-1/3}.\ear

Coherence effects between the emission from quarks $q_1$ and $q_2$
should
certainly show up if the integration length (\ref{A.18}) exceeds the
time
interval $\Delta t_0$ (\ref{2.16}) for which $q_1$ and $q_2$ can be
considered
as moving in uncorrelated background fields:
\bear\label{A.19}
\Delta t&\grgl& \Delta t_0,\nonumber\\
\Longrightarrow \quad
\omega&\klgl&m_q\left(\frac{m_q}{\sigma}\right)^2\left(\frac{l_{\rm
eff}}{R}
\right)^2\frac{a}{R}=:\overline\omega\ear
With $m_q=10\ {\rm MeV},\ \sigma=300\ {\rm MeV}, \ a=0.35\ \fm,\ R=1\
\fm$, and
$l_{\rm eff}=30\ \fm$ this leads to
\be\label{A.20}
\overline\omega=3.5\ {\rm MeV}.\ee

For $\omega\grgl\overline\omega$ on the other hand we have $\Delta
t\klgl\Delta
t_0$ and the almost wee quarks $q_1$ and $q_2$ travel during the
emission in
uncorrelated colour domains. For ``harder'' quarks $q_1$, $q_2$ this
is true a
forteriori since $\Delta t(x_1)$ defined in (\ref{2.14}) grows
proportional to
$\gamma$ (cf.~(\ref{2.13c})) whereas $\Delta t$ of (\ref{A.15}) is
only
proportional to $\gamma^{2/3}$.
The emission rate (\ref{A.13}) is determined by
\bear\label{A.21}
|\mbox{\boldmath{$\varepsilon$}}^*\vec Z(\vec
k)|^2&=&|\mbox{\boldmath{$\varepsilon$}}^*\vec Z_1(\vec
k)+\mbox{\boldmath{$\varepsilon$}}^*\vec Z_2(\vec k)|^2\nonumber\\
&=&|\mbox{\boldmath{$\varepsilon$}}^*\vec Z_1(\vec
k)|^2+|\mbox{\boldmath{$\varepsilon$}}^*\vec Z_2(\vec k)|^2
+2\mbox{Re}\left[(\mbox{\boldmath{$\varepsilon$}}^*\vec Z_1(\vec
k))(\mbox{\boldmath{$\varepsilon$}}\cdot\vec Z^*_2(\vec
k))\right].\ear
Looking at (\ref{A.12}) we see that the 1--2 interference term on the
r.h.s. of
(\ref{A.21}) is proportional to $\mbox{\boldmath{$\varepsilon$}}^*
\dot{\vec z}_{1,0}\,,\ \mbox{\boldmath{$\varepsilon$}}^*
\ddot{\vec{z}}_{1,0}$ multiplied  by
$\mbox{\boldmath{$\varepsilon$}}\dot{\vec{z}}_{2,0}\,,\
\mbox{\boldmath{$\varepsilon$}}\ddot{\vec{z}}_{2,0}$.
But in the case we are considering these transverse (with respect to
$\vec k$)
velocities and accelerations of the quarks  $q_1,q_2$ at the times
$t_{1,0},t_{2,0}$ are uncorrelated. In a sum over histories which we
have to
perform in (\ref{A.21}) the interference term should, therefore,
vanish.

To summarize: In this appendix we have given arguments that for
$\omega\grgl\overline\omega$ (\ref{A.20}) the emission of synchrotron
photons from quarks in a hadron can be added incoherently while  for
$\omega\klgl\overline\omega$ coherence effects will be important. We
have
worked here in a fixed reference frame where the  hadron and its
quarks move
reasonably fast. The angles of the quark velocity vectors relative to
the
hadron's velocity vector will vary from zero up to angles of the
order
$\sigma/E^{\rm max}_{\rm wee}\simeq\sigma\, a={\cal O}(1)$. The
emission angles
of the synchrotron photons with respect to the hadron's  velocity
vary then also up to angles of order 1 and the coherence condition
$\omega\klgl\overline\omega$ implies then
\be\label{A.22}
|\vec k_T|\klgl\overline\omega\ee
where $\vec k_T$ is transverse with respect to the hadron's velocity
vector.

To obtain the results from \cite{2} we still have to average over
all suitable
reference frames obtained by a boost  along the  collision axis from
the c.m.
system.  Since $\vec k_T$ is invariant under such boosts we would
then
expect from (\ref{A.22}) to see coherence for  photon transverse
(with
respect to the collision axis) momenta:
\be\label{A.23}
|\vec k_T|\klgl\overline\omega\ee
and incoherence for
\be\label{A.24}
|\vec k_T|\grgl\overline\omega.\ee
Of course, (\ref{A.23}), (\ref{A.24}) should be considered only as
\underbar{rough} guides for the regions of coherence and incoherence.

\section*{Appendix B}
\renewcommand{\theequation}{B.\arabic{equation}}
\setcounter{equation}{0}

In this appendix we discuss a simple model in order to make clear
what is
meant by ``dressed'' and ``naked'' hadrons and by the symbolic
equations
(\ref{2.29}) and (\ref{2.30}).

Consider a model with two scalar hermitean fields: a field $\psi(x)$
of
mass $m$ and a zero mass field $\varphi(x)$. As Lagrangian we choose:
\bear\label{B.1}
L(t)&=&\int d^3x\frac{1}{2}\left\lbrace
\partial_\mu\psi(x)\partial^\mu\psi
(x)-m^2\psi^2(x)+\partial_\mu\varphi(x)\partial^\mu\varphi(x)\right.
\nonumber\\
&&\phantom{\int d^3x\frac{1}{2}}\left.+2g\psi^2(x)\varphi(x)
-2\psi^2(x)V(\vec x)\right\rbrace,\qquad x=(t,\vec x).\ear
The field $\psi$ is the analogue of the quark and hadron fields,
the static potential $V(\vec x)$ represents the strong interaction,
$\varphi$ and $g$ are the analogues of the photon field and the
electromagnetic coupling, respectively.

We want to study the reactions
\bear\label{B.2}
({\rm a}) && \psi(\vec p)\to \psi(\vec p '),\nonumber\\
({\rm b}) && \psi(\vec p)\to \psi(\vec p ')+\varphi(\vec k),\ear
where for simplicity we work only up to first order in the coupling
constant $g$ and the potential $V$.

The Feynman rules following from (\ref{B.1}) are shown in Fig. 11,
where
\be\label{B.3}
\tilde V(\vec q):=\int d^3x\;e^{-i\vec q\vec x}V(\vec x).\ee
The diagrams for the reactions (a), (b) of (\ref{B.2}) are shown in
Fig. 12. They lead to the following S-matrix elements:
\be\label{B.4}
\langle\psi(\vec p ')|S|\psi(\vec p)\rangle=-4i\pi\delta({p'}^0-p^0)
\tilde V(\vec p '-\vec p),\ee
\be\label{B.5}
\langle\psi(\vec p ')\varphi(\vec k)|S|\psi(\vec p)\rangle=
8\pi ig\delta({p'}^0+\omega-p^0)\tilde V(\vec p '+\vec k-\vec p)
\left[-\frac{1}{2pk}+\frac{1}{2p'k}\right],\ee
where $\omega\equiv k^0$.
The cross sections for reactions (a), (b) of (\ref{B.2}) are
\be
\label{B.5a}
\frac{d\sigma^{(a)}}{d\Omega'}=\frac{1}{16\pi^2}|2\tilde V(\vec
q)|^2,\ee
\be\label{B.5b}
\frac{\omega d\sigma^{(b)}}{d\Omega'd^3k}=\frac{|\vec p '|}{|\vec p|
\,16\pi^2}|2\tilde V(\vec q+\vec k)|^2
\frac{2g^2}{(2\pi)^3}\left|\frac{1}{2pk}-\frac{1}{2p'k}\right|^2,
\ee
where $\vec q:=\vec p '-\vec p$ and $d \Omega'$ is the solid angle
element corresponding to $\vec p '$. All this is, of course,
completely
standard.

Now we will rederive these formulae using ``old-fashioned'' methods,
i.e. in
the Hamiltonian approach (cf.~e.g.~\cite{43}). From the Lagrangian
(\ref{B.1})
we find the canonical momenta conjugate to $\psi(x)$ and $\varphi(x)$
as
\bear\label{B.6}
\Pi_\psi(x)&=&\dot{\psi}(x),\nonumber\\
\Pi_\varphi(x)&=&\dot{\varphi}(x)\ear
and the Hamiltonian
\be\label{B.7}
H(t)=H_0(t)+H_g(t)+H'(t)\ee
where
\bear\label{B.8}
&&H_0(t)=\int d^3x\frac{1}{2}\left[\left(\Pi_\psi(x)\right)^2+
\left(\mbox{\boldmath$\nabla$}\psi(x)\right)^2
+m^2\psi^2(x)+\left(\Pi_\varphi(x)\right)^2+(\mbox{\boldmath$\nabla$}
 \varphi(x))^2
\right],
\nonumber\\
&&H_g(t)=-g\int d^3x\;\psi^2(x)\varphi(x),\nonumber\\
&&H'(t)=\int d^3x\;\psi^2(x)V(\vec x).\ear
We work now in the Schr\"odinger representation, setting $t=0$ in the
dynamical
variables and in the Hamiltonian,
\bear\label{B.9}
\psi(\vec x)&\equiv&\psi(0,\vec x),\nonumber\\
H&\equiv&H(0),\qquad
{\rm etc.}\ear
We then impose the canonical commutation relations for
$\psi,\Pi_\psi$
and $\varphi,\Pi_\varphi$ and expand these field operators in terms
of
annihilation and creation operators (cf. chapter 3 of \cite{13} for
all
conventions concerning normalizations). Let us denote by $b(\vec p)$
$(b^\dagger(\vec p) )$ the annihilation (creation) operators for
$\psi,\Pi_\psi$ and by $a(\vec k)$
$(a^\dagger(\vec k))$
those for $\varphi,\Pi_\varphi$. The next step is to expand
$H_0,H_g,H'$ in
terms of these operators and to normal order them. We get:
\be\label{B.10}
H_0=\int\frac{d^3k}{(2\pi)^32\omega}\,\omega\, a^\dagger(\vec
k)a(\vec k)
+
\int\frac{d^3p}{(2\pi)^32p^0}\,p^0\,b^\dagger(\vec p)b(\vec p),
\qquad(\omega\equiv k^0),
\ee
\bear\label{B.11}
H_g&=&-g\int\frac{d^3\,p'd^3p\,d^3k}{(2\pi)^6
2{p'}^02p^02\omega}\nonumber\\
&&\left\lbrace b^\dagger(\vec p ')b^\dagger(\vec p)\left[
\delta^{(3)}(\vec p '+\vec p+\vec k)a^\dagger(\vec
k)+\delta^{(3)}(\vec p '
+\vec p-\vec k)a(\vec k)\right]\right.\nonumber\\
&&+2 b^\dagger(\vec p ')b(\vec p)\left[
\delta^{(3)}(\vec p '-\vec p+\vec k)a^\dagger(\vec
k)+\delta^{(3)}(\vec p '
-\vec p-\vec k)a(\vec k)\right]\nonumber\\
&&\left.+b(\vec p ')b(\vec p)\left[
\delta^{(3)}(\vec p '+\vec p-\vec k)a^\dagger(\vec
k)+\delta^{(3)}(\vec p '
+\vec p+\vec k)a(\vec k)\right]\right\rbrace,
\ear
\bear\label{B.12}
H'&=&\int\frac{d^3p'd^3p}{(2\pi)^6 2{p'}^0 2p^0}
\left\lbrace b^\dagger(\vec p ')b^\dagger(\vec p)\tilde V(\vec p
'+\vec p)
\right.\nonumber\\
&&
\left.+2b^\dagger(\vec p ')b(\vec p)\tilde V(\vec p '- \vec p)
+b(\vec p ')b(\vec p)\tilde V(-\vec p '-\vec p)\right\rbrace .
\ear
We define now the ``naked'' states to be the eigenstates of $H_0$ and
denote
them by round brackets $|\phantom{0})$:

Naked vacuum: $|0)$
\bear\label{B.13}
a(\vec k)|0)&=&0,\nonumber\\
b(\vec p)|0)&=&0,\ear
for all $\vec k, \vec p$. It follows:
\be\label{B.14}
H_0|0)=0.\ee
Naked one-particle states:
\bear\label{B.15}
|\psi(\vec p))&=&b^\dagger(\vec p)|0),\nonumber\\
|\varphi(\vec k))&=&a^\dagger(\vec k)|0);\ear
\bear\label{B.16}
H_0|\psi(\vec p))&=&p^0|\psi(\vec p)),\nonumber\\
H_0|\varphi(\vec k))&=&\omega|\varphi(\vec k)).
\ear

Our next task is to construct the eigenstates of $H_0+H_g$. Denoting
these states by the brackets $|\phantom{0}\rangle$ we get in
perturbation
theory
up to first order in $g$ for the one-$\psi$-particle states:
\be\label{B.17}
|\psi(\vec p)\rangle=|\psi(\vec p))+g|\psi\varphi;\vec p)+
g|\psi^3\varphi;\vec p),\ee
where
\bear\label{B.18}
|\psi\varphi;\vec
p)&=&2\int\frac{d^3k}{(2\pi)^32\omega}\frac{1}{2p^0_1}
\frac{1}{p^0_1+\omega-p^0}|\psi(\vec p_1)\varphi(\vec k)),\nonumber\\
&&\vec p_1=\vec p-\vec k,\ear
and $|\psi^3\varphi;\vec p)$ is a state with 3 naked $\psi$ and one
naked $\varphi$ particles. The ``cloud'' of naked $\varphi$
particles in  a ``physical'' $\psi$ state is given by the
admixtures $|\psi\varphi;\vec p)$ and
$|\psi^3\varphi;\vec p)$ in (\ref{B.17}). The $\varphi$-particle
density is defined as
\be\label{B.19}
\frac{\omega dn(\vec k;\vec p)}{d^3k}=\frac{\langle\psi(\vec
p)|\frac{\textstyle a^\dagger(\vec k)a(\vec k)}{\textstyle 2
(2\pi)^3}|\psi(\vec p)\rangle}
{\langle\psi(\vec p)|\psi(\vec p)\rangle}\,.
\ee

In the following we will  consider the high energy limit in reactions
(a), (b) of (\ref{B.2}), to be precise we require
\be\label{B.20}
{p'}^0,p^0\gg m,\omega.\ee
In this limit the contribution $|\psi^3\varphi;\vec p)$ in
(\ref{B.17})
can be neglected. It leads to a $Z$-diagram in the reaction (b) of
(\ref{B.2})
being suppressed by $1/p^0$ relative to the leading term. We get then
for the
density (\ref{B.19})
\be\label{B.21}
\frac{\omega dn(\vec k;\vec p)}{d^3k}\simeq
\frac{2g^2}{(2\pi)^3}\left|
\frac{1}{2(pk)}\right|^2.\ee

Consider now the following scattering reaction induced by the
Hamiltonian
$H'$ (\ref{B.12}):
\be\label{B.22}
|i\rangle\longrightarrow |f\rangle\ee
where $|i\rangle$ and $|f\rangle$ are eigenstates of $H_0+H_g$ with
energies $E_i,E_f$. In first Born approximation we have
\be\label{B.23}
\langle f|S|i\rangle=\delta_{fi}-2\pi i\delta(E_f-E_i)\langle
f|H'|i\rangle.
\ee
For the reactions (\ref{B.2}) we have to take for $|i\rangle$ the
state
$|\psi(\vec p)\rangle$ (\ref{B.17}). Applying $H'$ (\ref{B.12}) to it
we get:
\bear\label{B.24}
H'|\psi(\vec p)\rangle&=&\int\frac{d^3p_2}{(2\pi)^32p^0_2}\,2\,
\tilde V(\vec
p_2
-\vec p)|\psi(\vec p_2))\nonumber\\
&&+2g\int\frac{d^3p_2d^3k}{(2\pi)^6 2p^0_2\, 2\omega}\frac{2\tilde
V(\vec
p_2+\vec k-\vec p)}{2p^0_1(p^0_1+\omega-p^0)}|\psi(\vec
p_2)\varphi(\vec k))
+...\ear
Here the dots stand for terms which do not contribute to the
scattering
reactions (\ref{B.2}) due to energy conservation and/or particle
content
and for terms which give vanishing contribution in the limit
(\ref{B.20}).
The interpretation of (\ref{B.24}) is clear. The action of $H'$
scatters the
naked $\psi$ particle out of the physical one and produces a
superposition
of states with one naked $\psi$ particle $\left(|\psi(\vec
p_2))\right)$
and one naked $\psi$ and one $\varphi$ particle $\left(|\psi(\vec
p_2)
\varphi(\vec k))\right)$. Now we reexpress these naked particle
states
in terms of physical ones. Inverting (\ref{B.17}) we find to order
$g$:
\be\label{B.25}
|\psi(\vec p_2))=|\psi(\vec p_2)\rangle-g|\psi\varphi;\vec p_2\rangle
-g|\psi^3\varphi;\vec p_2\rangle.\ee
Inserting this in (\ref{B.24}) and neglecting again all terms which
vanish
in the limit (\ref{B.20}) leads to
\bear\label{B.26}
H'|\psi(\vec p)\rangle&=&\int\frac{d^3p_2}{(2\pi)^3 2p^0_2}2\tilde V
(\vec p_2-\vec p)|\psi(\vec p_2)\rangle-g\int\frac{d^3p_2}{(2\pi)^3
2p^0_2}2\tilde V
(\vec p_2-\vec p)|\psi\varphi;\vec p_2\rangle\nonumber\\
&&+2g\int\frac{d^3p_2d^3k}{(2\pi)^6 2p^0_2\, 2\omega}
\frac{2\tilde V(\vec p_2+\vec k-\vec
p)}{2p^0_1(p^0_1+\omega-p^0)}|\psi(\vec
p_2)\varphi(\vec k)\rangle
+... .\ear
The $S$-matrix elements for the reactions (a), (b) of (\ref{B.2}) are
now
easily obtained from (\ref{B.23}) and (\ref{B.26}).
For reaction (a) of (\ref{B.2}) only the first term on the r.h.s.
of (\ref{B.26}) contributes and gives for the $S$-matrix element
just the expression in (\ref{B.4}). For reaction (b) of (\ref{B.2})
we
get two contributions in the $S$-matrix element:
\be\label{B.27}
\langle\psi(\vec p ')\varphi(\vec k)|S|\psi(\vec p)\rangle
=-2\pi i\delta({p'}^0+\omega-p^0)\left[A(\vec p',\vec k;\vec
p)+A'(\vec p ',
\vec k;\vec p)\right],\ee
where
\bear\label{B.28}
A(\vec p',\vec k;\vec p)&=&\frac{4g\tilde V(\vec p'+\vec k-\vec p)}
{2p^0_1(p^0_1+\omega-p^0)}\simeq\frac{4g
\tilde V(\vec p'+\vec k-\vec p)}{2pk}\, ,
\nonumber\\
A'(\vec p',\vec k;\vec p)&=&-\frac{4g\tilde V(\vec p'+\vec k-\vec p)}
{2{p'}^0_1({p'}^0+\omega-{p'}_1^0)}\simeq-\frac{4g
\tilde V(\vec p'+\vec k-\vec p)}{2p_1k}\, ,
\nonumber\\&&
\vec p'_1=\vec p'+\vec k.\ear
The amplitude $A$ in (\ref{B.27}) arises from the third term on the
r.h.s. of (\ref{B.26}). It represents the emission of $\varphi$
particles
due to the kicking out of the naked $\psi$ particle form the original
state
$|\psi\rangle$ in the scattering process. The amplitude $A'$ in
(\ref{B.27})
arises from the second term on the r.h.s. of (\ref{B.26}). It
represents
the emission of $\varphi$-particles due to the ``dressing'' of the
kicked-out naked $\psi$ particle in the final state.

In the high energy limit (\ref{B.20}) the expression for the
$S$-matrix
element (\ref{B.27}) goes over into (\ref{B.5}), derived from the
Feynman rules. Of course, we get exactly the Feynman rules result
with
the Hamiltonian method if we do not make the high energy
approximation.

Let us now consider a potential $V(\vec x)$ of extension $\Delta x$.
We expect then the typical momentum transfer $\vec q$ to be of order
\be\label{B.29}
|\vec q|\sim\frac{1}{\Delta x}.\ee
Furthermore we require $\vec k$ to be  such that  in the cross
section
(\ref{B.5b}) the interference term between the two terms $(2pk)^{-1}$
and
$(2p'k)^{-1}$, i.e. the two amplitudes $A,A'$ of
(\ref{B.27}) is negligible and that
\be\label{B.30}
\tilde V(\vec q+\vec k)\simeq\tilde V(\vec q).\ee
This requires
\bear\label{B.31}
|\vec q|^2&\gg& m^2+\omega^2,\nonumber\\
 \omega&\ll&\frac{1}{|\vec q|\Delta x^2}.\ear
We get then for the cross section of reaction (b) of (\ref{B.2}) in
the limit (\ref{B.20}):
\be\label{B.32}
\frac{\omega\,
d\sigma^{(b)}}{d\Omega'd^3k}\simeq\frac{d\sigma^{(a)}}{d\Omega'}
\left\lbrace\frac{\omega\, dn(\vec k;\vec p)}{d^3k}+\frac{\omega\,
dn(\vec
k;\vec p ')}{d^3k}\right\rbrace.\ee
The total number density of $\varphi$ particles produced in the
reaction $(b)$
of (\ref{B.2})
\be\label{B.33}
\frac{\omega\, dn_{\rm tot}(\vec k)}{d^3k}:=\frac{\omega
\,d\sigma^{(b)}}{d\Omega'd^3k}
\bigg/ \frac{d\sigma^{(a)}}{d\Omega'}\ee
is thus given by the sum of the densities of ``naked'' $\varphi$
particles in
the original and in the
scattered $\psi$ particles. We have made the analogous assumption
for our synchrotron effect in (\ref{2.33}), (\ref{2.34}). Of
course, the number densities (\ref{B.21}), (\ref{B.33}) in our model
are proportional to $\omega^{-2}$ corresponding to
bremsstrahlung, whereas for the synchrotron effect we derived a
behaviour proportional to $\omega^{-4/3}$. But  this
should  not spoil the general argument. For sake of orientation,
let us finally use the estimates (\ref{B.31}) for the case of
hadron-hadron
scattering. Then we would identify $\psi$ with the quark field and
set
\be\label{B.34} \Delta x\simeq a\ee
where $a$ is the correlation length (\ref{2.4}). The typical momentum
transfers should then satisfy
\be\label{B.35}
|\vec q|\klgl 1/a= 570\ {\rm MeV}\ee
and from (\ref{B.31}) we find that for (\ref{B.32}) to hold
we should require for $\omega$:
\be\label{B.36}
\omega^2+m^2_q\ll|\vec q|^2,\ee
\be\label{B.37}
\omega\ll\frac{1}{|\vec q|\Delta x^2}\simeq\frac{1}{a}=570\ {\rm
MeV}.
\ee
For light quarks with $m_q\simeq 10$ MeV both (\ref{B.36})
and (\ref{B.37}) are satisfied for $\omega\klgl
200$ MeV which is very reasonable and agrees with the estimates which
were
made in \cite{2}.

To conclude: In this appendix we have studied a model where we could
give
a precise meaning to the ``$\varphi$ cloud'' of a particle, the
analogue
of the ``$\gamma$ cloud'' of a hadron, and to analogues of
(\ref{2.29}),
(\ref{2.30}), (\ref{2.33}), (\ref{2.34}).

\newpage
\section*{Figure Captions}
\begin{description}
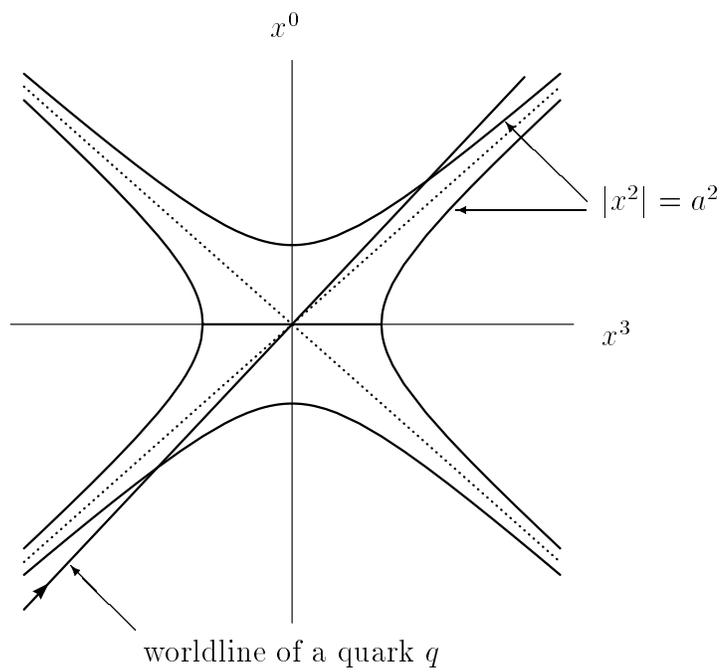
\item[Fig. 1:] \quad Sketch of a ``colour domain'' in Minkowski space
and of a quark from a fast hadron moving through it.
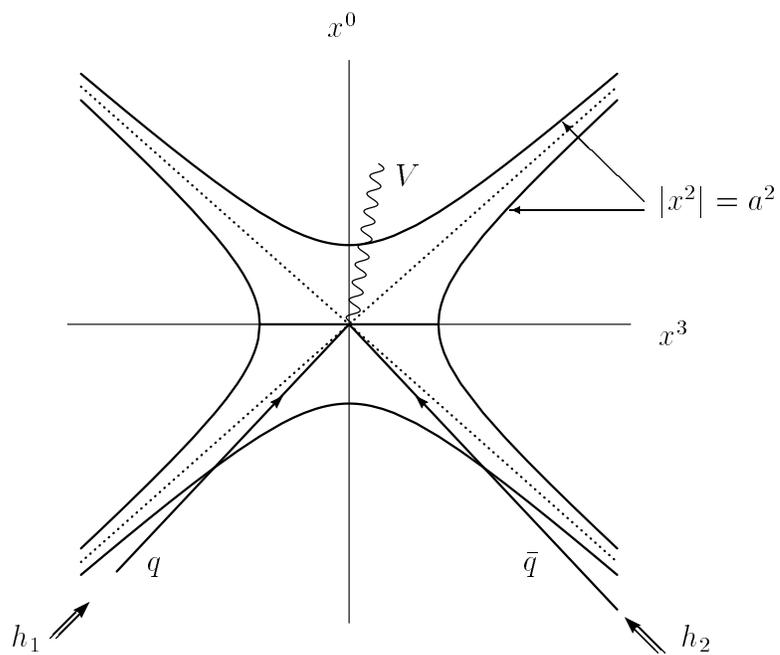
\item[Fig. 2:] \quad Annihilation of a $q\bar q$ pair and production
of
a vector boson $V=\gamma^*, Z, W$ in a colour domain. Here $q$ and
$\bar q$
come from two different hadrons $h_1$ and $h_2$, respectively.
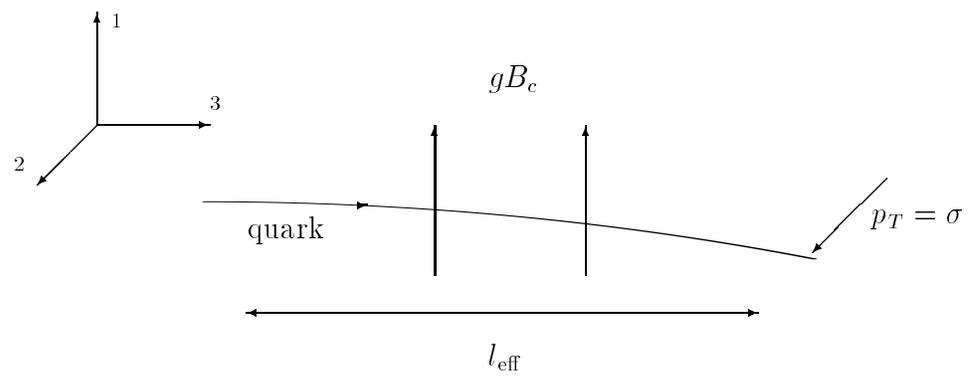
\item[Fig. 3:] \quad A quark moving in 3-direction in a transverse
(in
1-direction) chromomagnetic field of strength $gB_c$ and picking up a
transverse momentum (in 2-direction) of magnitude $\sigma$ over a
length
$l_{\rm eff}$. Here $\sigma$ is the mean transverse momentum of
quarks in
the hadron.
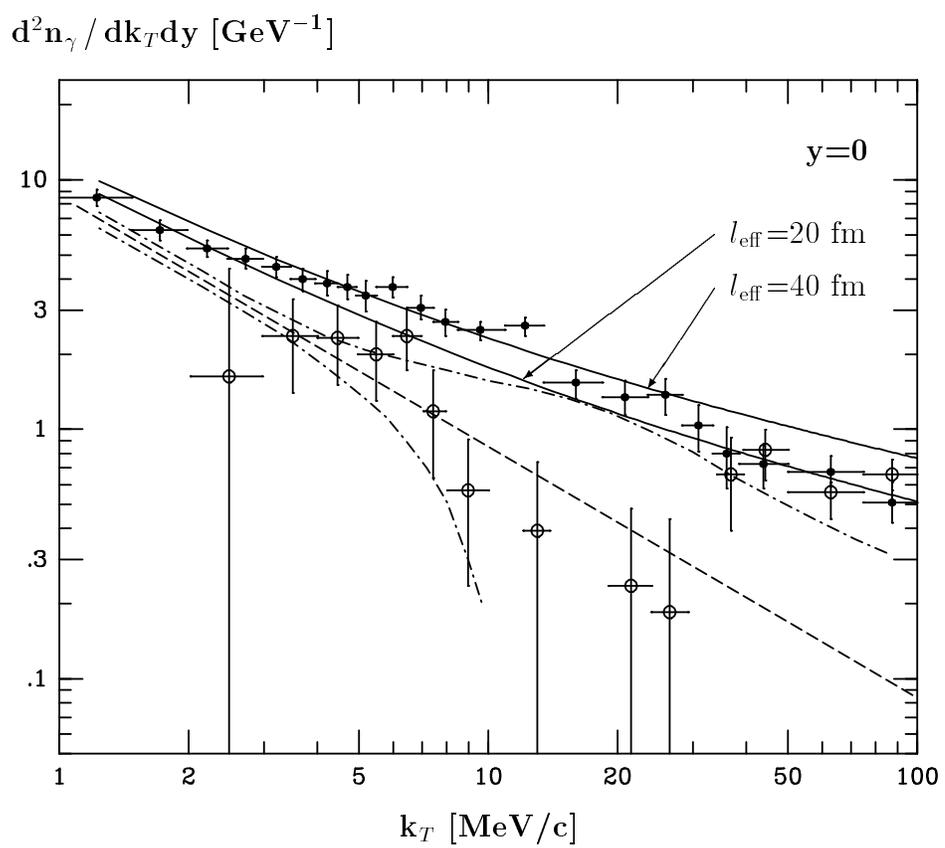
\item[Fig. 4:] \quad
The $|{\bm k}_T|$ distribution for direct photons emitted at c.m.\
rapidity
$y=0$ in $p$--Be collisions at 450 GeV incident proton momentum from
\cite{3}.
The normalization is according to a private communication by H. J.
Specht. The
background of decay photons is subtracted. The dash--dotted line
gives the
expected yield of photons from hadronic bremsstrahlung, the dashed
lines show
the upper and lower limits including the systematic errors in the
shape of the
decay background and the bremsstrahlung calculation (cf.\ \cite{3}).
The lower
(upper) solid line is the
result of our calculation for synchrotron photons ((\ref{3.4}) ff.)
with
$l_{\rm eff}=20$ fm ($l_{\rm eff}=40$ fm) added to the spectrum of
hadronic
bremsstrahlung.
\item[Fig. 5:] \quad
The structure functions $F_2^p(x),F_2^N(x)$ ($N$=average of proton
and
neutron) and $F_2^\pi(x)$ for $Q^2=4{\GeV}^2$ used in this paper.
$F_2^p,F_2^N$ are from set $D_0'$ of \cite{24}, $F_2^\pi$ is from
\cite{27}.
\item[Fig. 6:] \quad The quantities $2S^{(p)}(\cos\vartheta^*)$ and
$\Sigma(\cos\vartheta
^*)$ (cf. (\ref{3.5})) related to the angular distribution of
synchrotron photons in the overall c.m. frame according to
(\ref{3.4}).
Here $\vartheta^*$ is the emission angle of the photon:
$\cos\vartheta^*=\hat{\bm{p}}_1\cdot \hat{\bm{k}}$.
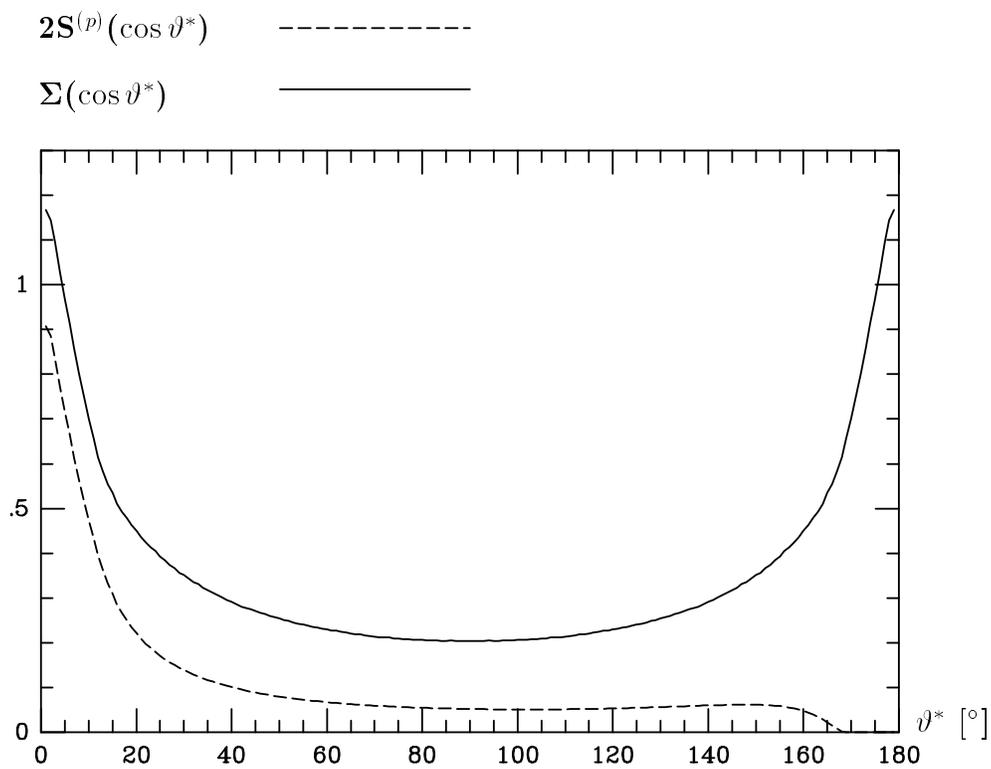
\item[Fig. 7:] \quad The quantity $(\sin\vartheta^*)^{2/3}\cdot
\Sigma(\cos\vartheta^*)$ as
function of $\vartheta^*$. Over a large range of $\vartheta^*$ this
quantity
is nearly constant.
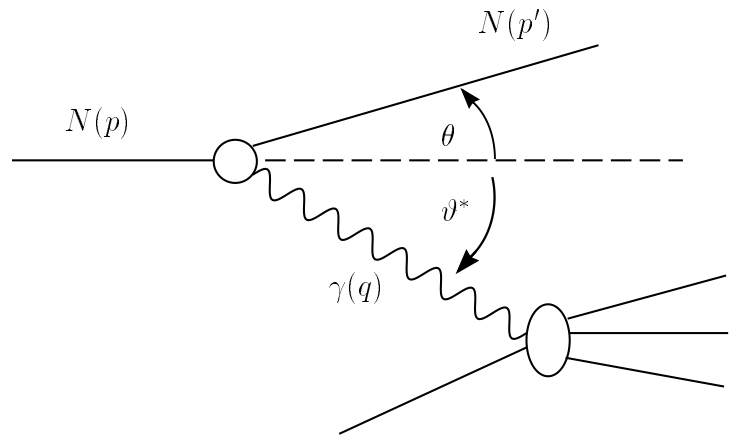
\item[Fig. 8:] \quad A nucleon interacting by emission of a
quasi-real photon.
\item[Fig. 9:] \quad The data for the electric formfactor of the
neutron,
$G_E^n(Q^2)$ from refs. \cite{37,38}.
Full line: our ``synchrotron'' prediction $\propto(Q^2)^{1/6}$
normalized
to the data at $Q^2=5\ \fm^{-2}$.
Dashed line: the slope of $G_E^n(Q^2)$ at $Q^2=0$ as deduced from
thermal
neutron-electron scattering \cite{39}.
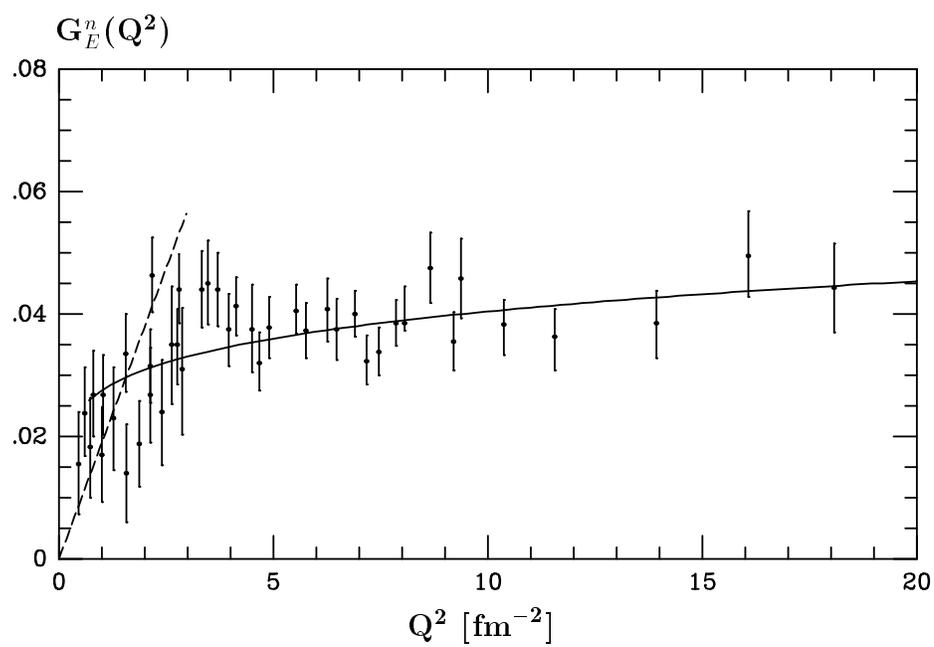
\item[Fig. 10:] \quad Same as Fig.~9 but in a double logarithmic
plot.
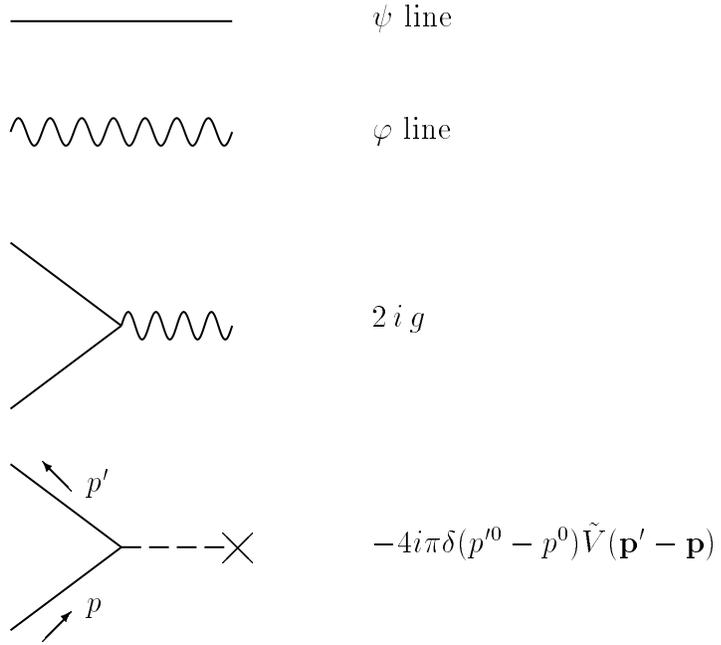
\item[Fig. 11:] \quad Feyman rules following from the Lagrangian
(\ref{B.1}).
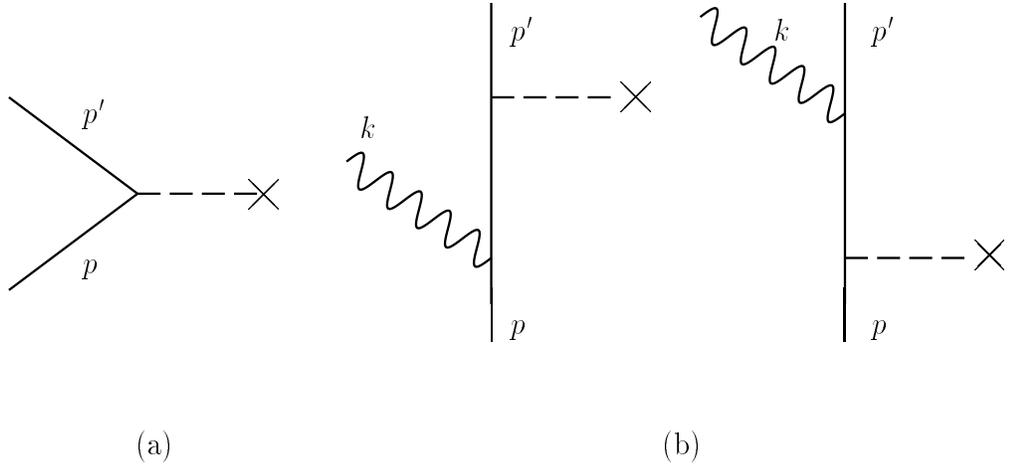
\item[Fig. 12:] \quad Lowest order diagrams for the reactions (a) and
(b) of
(\ref{B.2}).
\end{description}

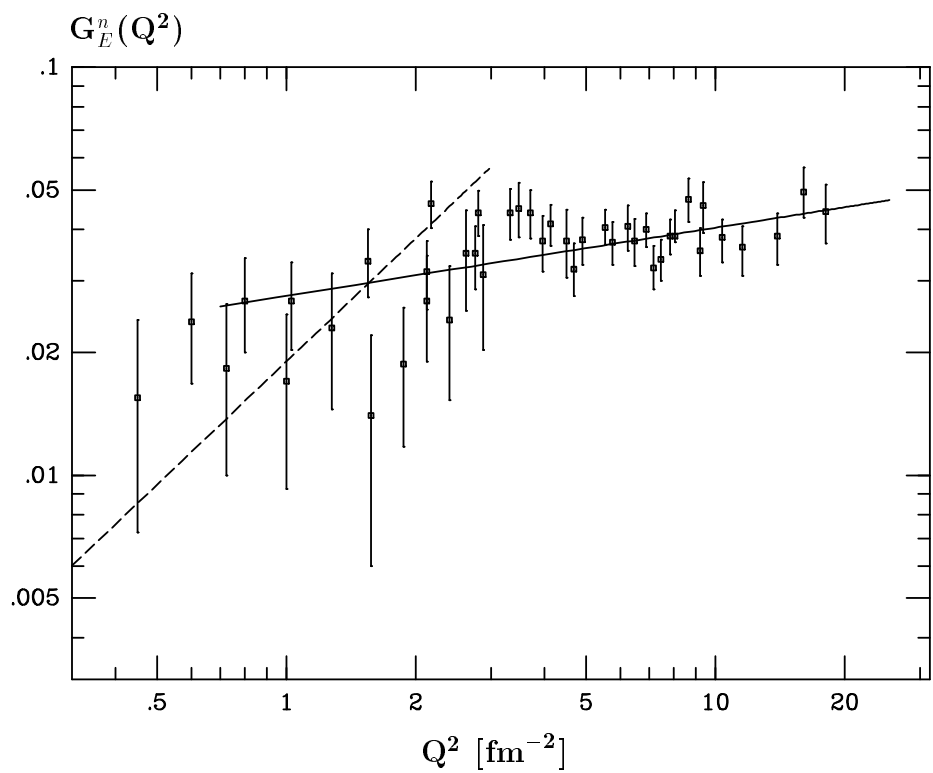
\end{document}